\documentclass[prd,nofootinbib,english]{revtex4}
\usepackage{amsmath}
\usepackage{amsfonts,color}
\usepackage{amssymb,float}
\usepackage[utf8]{inputenc}
\allowdisplaybreaks
\usepackage{color}
\usepackage{accents}
 \usepackage{hyperref}
\hypersetup{
    colorlinks=true,
    linkcolor=blue,
    filecolor=magenta,      
   citecolor=blue
}

\usepackage{enumitem,wasysym,textcomp}

\usepackage{tikz}
\usetikzlibrary{shapes.geometric}
\usetikzlibrary{arrows.meta,arrows}

\newcommand{\udt}[3]{#1^{#2}_{\phantom{#2}#3}}
\newcommand{\udut}[4]{#1^{#2\phantom{#3}#4}_{\phantom{#2}#3\phantom{#4}}}

\newcommand{\dut}[3]{#1_{#2}^{\phantom{#2}#3}}
\newcommand{\dudt}[4]{#1_{#2\phantom{#3}#4}^{\phantom{#2}#3}}

\newcommand{\lc}[1]{\mathring{#1}}

\begin{document}
\title{Solar System Tests in Modified Teleparallel Gravity}

\author{Sebastian Bahamonde}
\email{sbahamonde@ut.ee, sebastian.beltran.14@ucl.ac.uk}
\affiliation{Laboratory of Theoretical Physics, Institute of Physics, University of Tartu, W. Ostwaldi 1, 50411 Tartu, Estonia}
\affiliation{Laboratory for Theoretical Cosmology, Tomsk State University of
Control Systems and Radioelectronics, 634050 Tomsk, Russia (TUSUR)}

\author{Jackson Levi Said}
\email{jackson.said@um.edu.mt}
\affiliation{Institute of Space Sciences and Astronomy, University of Malta, Malta}
\affiliation{Department of Physics, University of Malta, Malta}

\author{M. Zubair}
\email{mzubairkk@gmail.com; drmzubair@cuilahore.edu.pk}
\affiliation{Department of Mathematics, COMSATS University Islamabad, Lahore Campus, Lahore-Pakistan}

\begin{abstract}
In this paper, we study different Solar System tests in a modified Teleparallel gravity theory based on an arbitrary function $f(T,B)$ which depends on the scalar torsion $T$ and the boundary term $B$. To do this, we first find new perturbed spherically symmetric solutions around Schwarzschild for different power-law forms of the arbitrary Lagrangian. Then, for each model we calculated the photon sphere, perihelion shift, deflection of light, Cassini experiment, Shapiro delay and the gravitational redshift. Finally, we confront these computations with different known experiments from these Solar System tests to put different bounds on the mentioned models. We then conclude that $f(T,B)$ is compatible with these Solar System experiments with a wide range of parameters which are relevant for cosmology.
\end{abstract}

\maketitle

\section{Introduction}
The study of properties of a gravitational theory on Solar System scales and investigations about the consistency of its cosmologically
viable models with Solar System bounds is an essential ingredient to establish bounds on the theory for future explorations. All gravitational effects in the Solar System are well understood in Einstein's Theory of General Relativity (GR) \cite{will_2018} and viable modified theories must confront them \cite{Clifton:2011jh,Capozziello:2011et}. These effects in the framework of Schwarzschild-de Sitter spacetime have been calculated in Ref.~\cite{Kagramanova:2006ax}, where the cosmological constant $\Lambda$ in the metric is considered as a free parameter. Here, it was measured how Solar System effects like gravitational redshift, light deflection, gravitational time delay, geodetic or de-Sitter precession, would be affected in the presence of the cosmological constant. Its impact on a Doppler measurement has also been taken into account and used to approximate the velocity of the Pioneer 10 and 11 spacecraft. Investigations have indicated that a cosmological constant with the value of $\Lambda_0\sim 10^{-52} m^{-2}$ becomes irrelevant for all these tests, which led to the conclusion that cosmic acceleration has no connection with observable Solar System effects. However, the value of $\Lambda \sim -10^{-37} m^{-2}$ is not found to be compatible with the observed perihelion shift. It was the first effort towards the general target of securing Solar System constraints for modified theories of gravity. Then, the $f(\lc{R})$ theories of gravity are investigated in Ref.~\cite{Ruggiero:2006qv} to see their effects on Solar System tests and these results are used to constrain the cosmological constant and particularly the $f(\lc{R})$ functions (here and throughout over-circles are used to denote quantities that are calculated using the Levi-Civita connection). In Ref.~\cite{Chiba:2006jp}, Solar System bounds are discussed for the general $f(\lc{R})$ gravity. There is another class of modified gravity models, i.e., so-called $f(\lc{G})$ theory of gravity where $\lc{G}$ is the standard gravity Gauss-Bonnet term. In Ref.~\cite{DeFelice:2009aj}, the authors focused on viable $f(\lc{G})$ models and declared their consistency with Solar System bounds against a wide range of parameters. The major reason found behind this consistency is that the Gauss-Bonnet scalar assumes a non-zero value $\lc{G} = 12r_s^2 /r^6$, where $r_s$ is the Schwarzschild radius, even in the vacuum spherically symmetric spacetime. The derived results are claimed to be suitable for application against any modified gravitational model which has power-law corrections to the Schwarzschild metric. In Ref.~\cite{Deng:2017grq}, improved Solar System bounds are obtained for this theory using cosmological models. These improved constraints are obtained on the behalf of supplementary advances of the perihelia given by INPOP10a and INPOP15a (France) and EPM2011 (Russia) ephemerides, where Lense-Thirring effect has been considered due to the Sun’s angular momentum and the uncertainty of the Sun’s quadrupole moment.

The teleparallel theory of gravity is built on teleparallel geometry in which the gravitational
source is provided by torsional formulation instead of curvature scalar structure of GR~\cite{Goenner2004,PhysRevD.19.3524,Arcos:2005ec,Maluf:2013gaa,Aldrovandi:2013wha}. We recall here the teleparallel theory is an idea of Einstein which is based on the concept of a flat connection but torsionful and with the tetrads being the fundamental object which is an orthogonal field based on the four-dimensional spacetime tangent space~\cite{Cai:2015emx,Hayashi:1979qx,Unzicker:2005in}. As the tetrad has sixteen components, Einstein called this structure as a unification of electromagnetism and gravity by relating six additional degrees of freedom to the electromagnetic field. Later, he realised that his idea was not correct. It turns out that it is possible to formulate an equivalent theory (in field equations) to GR but in this framework~\cite{Aldrovandi:2013wha,Maluf:2013gaa}, therefore, it is also termed as the teleparallel equivalent of GR (TEGR)~\cite{Moller:233632,Plebanski:1962xjr,PhysRevD.19.3524}. In this scenario, it turns out that those extra six degrees of freedom are related to the local Lorentz invariance of the theory rather than electromagnetism. It is considered one of the most interesting alternative to GR whose different torsion based extensions are proposed in literature to make it more favorable which include some modifications such as $f(T)$ theory of gravity is one of the popular models beyond the TEGR which is based on the replacement of torsion scalar $T$ in teleparallel gravity (TG) by an arbitrary function $f(T)$~\cite{Ferraro:2006jd,Bengochea:2008gz,Linder:2010py,Krssak:2015oua,DAgostino:2018ngy,Abedi:2018lkr}. 

Spherical and axial symmetries have not widely been explored yet in modified Teleparallel gravity and one of the reasons is the absence of physically viable exact solutions of field equations. In~Ref.~\cite{Hohmann:2019nat,Krssak:2015oua}, it was explained how to handle spherical symmetry in modified Teleparallel gravity. These solutions are based on the evaluation of the tetrad which needs to satisfy both the symmetric and antisymmetric parts of field equations together with the production of a torsion scalar which tends to zero for Minkowski spacetime limit. Some papers have found exact spherically symmetric solutions in $f(T)$ gravity, but the majority of them give rise to a constant torsion scalar which leads to the situation where $f(T)$ gravity corresponds to TEGR plus a cosmological constant. This has led to a number of inconsistent results which is only remedied when these anti-symmetric field equations are taken into account~\cite{Ferraro:2011ks,Paliathanasis:2014iva,Bejarano:2017akj}. In modified Teleparallel gravity, there are many incorrect or trivial solutions assuming $T= \textrm{conts.}$. Some studies assuming $T\neq \textrm{const.}$ are~\cite{Junior:2015dga,DeBenedictis:2018wkp,Daouda:2012nj,Aftergood:2014wla,Awad:2019jur,Bahamonde:2019jkf,Boehmer:2019uxv,Bahamonde:2016jqq}. Nevertheless, $f(T)$ gravity is still awaiting physically viable analytical solutions of field equations for spherical symmetry. 

The $f(T)$ theory of gravity has also been treated on Solar System  settings. In~Refs.\cite{Farrugia:2016xcw,Xie:2013vua,Ruggiero:2016iaq,Iorio:2015rla}, different Solar System effects which include perihelion precession, Shapiro time delay, gravitational redshift and light bending have been investigated in the context of $f(T)$ gravity. However, results obtained in these papers need to be corrected since they were computed using the incorrect perturbed solution~\cite{Farrugia:2016xcw,Xie:2013vua,Ruggiero:2016iaq,Iorio:2015rla} derived from the study~\cite{Ruggiero:2015oka}. In this paper, the authors used the incorrect field equations in $f(T)$ gravity. This can be easily seen since, if one sets $f(T)=T$ in Eqs.(8)-(10) in~\cite{Ruggiero:2015oka}, one does not recover GR. Moreover, if one uses the correct equations, one does not obtain any perturbed correction in power-law $f(T)$ around Minkowski (up to first order expansions).. In~\cite{DeBenedictis:2016aze} and later in \cite{Bahamonde:2019zea} the authors derived the photon sphere and the perihelion shift for weak $f(T)$ gravity and presented improvements and extensions for the already existing results. For this, the first order influence of a teleparallel power law $f(T )$ gravity perturbation of GR, has been presented in spherical symmetry and the power law perturbation of the type $f(T)=T+\epsilon (\alpha/2) T^p$ has been considered around two different geometries, i.e., Minkowski spacetime and Schwarzschild spacetime. The later case facilitated the calculation of photon sphere and the perihelion shift in $f(T)$ gravity. For the both impediments, it was found that the influence of the teleparallel perturbation is decreased with higher value of $p$, while the most strict bound for $\alpha$ is obtained against $p = 2$. 

 Another interesting approach to tackling Solar System tests is through the parametrized post-Newtonian formalism (PPN) which offers a platform on which to compare the wide range of theories GR. The PPN formalism produces ten parameters that can be compared with high precision Solar System data to establish the viable regions of a model and be used as an effective tool to characterize its Solar System behaviour. In TG, the PPN formalism has been studied in a variety of models, in Ref.~\cite{Ualikhanova:2019ygl} the a generalized framework in which $f(T)$ gravity is decomposed into the irreducible constituents of the torsion tensor is considered. In this work, they find that large parts of this general theory are identical making this largely indistinguishable within the Solar System. In many other works, the scenario of a coupled scalar field is considered such as Refs.~\cite{Flathmann:2019khc,Emtsova:2019qsl,Li:2013oef,Chen:2014qsa} with interesting results. Of particular mention is Ref.~\cite{Flathmann:2019khc} where a generalized approach is developed for the scalar field coupling. Another work along a similar vein is Ref.~\cite{Bahamonde:2020cfv} which considers the analog of standard Horndeski gravity within the TG context, giving a larger landscape of theories in which to form cosmological models. In this scenario it was found that for a large portion of the PPN parameter regions, the ensuing models are identical to GR and thus viable. The concept of coupling a scalar field with TG has also been studied in the context of scalar-tensor theories with the boundary term $B$ such as Ref.~\cite{Sadjadi:2016kwj}. However, this is a very particular case and the work does not incorporate the spin connection which makes their model local Lorentz violating (as will be explained further in the next section). Moreover, this model contains a coupling with a scalar field rather than a pure $f(T,B)$ theory. It is for this reason that we develop the precise Solar System tests in this work and study their constraints.

Different generalizations of $f(T)$ theory of gravity have also been offered in literature.
One of them  takes place by introducing a new Lagrangian scalar $f(T,B)$ which involves a boundary term
$B$ related to the divergence of the torsion tensor~\cite{Bahamonde:2015zma}. This theory becomes equivalent to $f(\lc{R})$ gravity for the choice of special form $f(-T + B)$ since the Einstein Hilbert action Lagrangian is dynamically equivalent to this argument choice. The latter is the only case in which Lorentz invariance can be achieved for a zero spin-connection irrespective of tetrad choice. Several studied about cosmology have been done in this theory~\cite{Bahamonde:2016grb,Paliathanasis:2017efk,Bahamonde:2016cul,Zubair2018,Paliathanasis:2017flf,Farrugia:2018gyz}. The traditional gravitomagnetic effects of the geodetic and Lense-Thirring phenomena have been explored and investigated in the context of $f(T,B)$ gravity, so that some viable models may be constrained within the theory~\cite{Farrugia:2020fcu}. Regarding, spherical symmetry, recently in~\cite{Bahamonde:2019jkf} some exact solutions were found but none of them are physically interesting.

In this manuscript we present the modified field equations for spherically symmetric spacetime in the framework of $f(T,B)$ theory and then we find new perturbed solutions around Schwarzschild for different power-law models of the arbitrary Lagrangian.  Then, we calculate different Solar System tests for the perturbed solutions. The Solar System is considered as a classical laboratory for testing the laws of gravity. Most important tests have been developed with the help of Solar System observations. These include tests based on the advance of the perihelion of Mercury, the Shapiro time delay to the Viking landers, the deflection of light as it passes near the Sun, the frequency shift of signals to the Cassini spacecraft, and the violation of the gravitational redshift found from a system composed from a spacecraft and the Earth.

This article is organized as follows: In Sec.~\ref{sec:tele}, a brief introduction of teleparallel gravity and its extensions is presented. Sec.~\ref{sec:sph} provides field equations in $f(T, B)$ gravity for spherical symmetry, which is followed by perturbed solutions for the symmetry generating various cases. Sec.~\ref{sec:solar} offers Solar System constraints and comprises of five subsections covering
photon sphere and perihelion shift, deflection of light, Cassini experiment and retardation of light
and, gravitational redshift and equivalence principle. Finally, in Sec.~\ref{sec:con}, we summarize the main results.

\section{Introduction to Teleparallel gravity and its Extensions}\label{sec:tele}
GR expresses gravitation through the metric tensor by means of the Levi-Civita connection, $\mathring{\Gamma}^{\sigma}_{\mu\nu}$, which is torsion-less \cite{misner1973gravitation} (over-circles are used throughout to represent quantities calculated using the Levi-Civita connection). Thus, it is through the connection that curvature is exhibited, and not through the metric itself. In this way, TG replaces the standard gravity connection with the Weitzenb\"{o}ck connection, $\Gamma^{\sigma}_{\mu\nu}$, which is curvature-less and satisfies the metricity condition \cite{Aldrovandi:2004db,Hayashi:1979qx}.

Through the substitution of the gravitational connection, the means by which gravitation is expressed can be altered from curvature to torsion. Now in GR, the Riemann tensor is extensively utilized since it gives a meaningful measure of curvature \cite{nakahara2003geometry}. The Riemann tensor also appears in many modified theories of gravity since they are built on GR \cite{Clifton:2011jh,Capozziello:2011et}. The Weitzenb\"{o}ck connection is curvature-less and thus its Riemann tensor components will always vanish irrespective of the metric components. It is for this reason that TG requires a complete reformulation of the contributing tensor quantities in order to build realistic models of gravity.

In GR, the metric tensor, $g_{\mu\nu}$, is the fundamental dynamical object of the theory, a property which permeates into many of its modifications \cite{Clifton:2011jh}. However, in TG the metric is a derived quantity that emerges from the tetrad, $\udt{e}{a}{\mu}$ \cite{Aldrovandi:2004db}. The tetrad connects the general manifold (Greek indices) to their tangent space analog (Latin indices), which renders the tetrad to be a soldering agent of the manifold in the theory \cite{Hehl:1994ue}. Therefore, the tetrads (and their inverses $\dut{e}{a}{\mu}$) can be used to transform between inertial and non-inertial indices through \cite{ortin2004gravity}
\begin{align}\label{metric_ten_def}
    g_{\mu\nu} = \udt{e}{a}{\mu}\udt{e}{b}{\nu}\eta_{ab}\,,& &\eta_{ab}=\dut{e}{a}{\mu}\dut{e}{b}{\nu}g_{\mu\nu}\,,
\end{align}
while also observing orthogonality conditions
\begin{align}
    \udt{e}{a}{\mu}\dut{e}{b}{\mu} = \delta_a^b\,,& &\udt{e}{a}{\mu}\dut{e}{a}{\nu} = \delta_{\mu}^{\nu}\,,
\end{align}
for consistency's sake. Using the tetrad, the Weitzenb\"{o}ck connection can then be defined as
\begin{equation}
    \Gamma^{\sigma}_{\mu\nu} := \dut{e}{a}{\mu}\partial_{\mu}\udt{e}{a}{\nu} + \dut{e}{a}{\sigma}\udt{\omega}{a}{b\mu}\udt{e}{b}{\nu}\,,
\end{equation}
where $\udt{\omega}{a}{b\mu}$ represents the spin connection. The Weitzenb\"{o}ck connection represents the most general linear affine connection that is both curvature-less and satisfies the metricity condition. The explicit appearance of the spin connection occurs to preserve the general covariance of the ensuing field equations of the theory \cite{Krssak:2015oua}. Theories based on the Levi-Civita connection also have nonvanishing spin connection components but these are hidden in the internal structure of the theory \cite{nakahara2003geometry}. In TG, the spin connection plays an active role in the field equations to retain invariance under local Lorentz transformations (LLTs), i.e. the freedom in the choice of inertial frames that permeates through the tetrads is accounted for in the values of the spin connection components. In this way, for any choice of spacetime, there exists a choice of tetrad that allows for zero spin connection components \cite{Maluf:2013gaa}.

The spin connection is thus a totally inertial quantity, and by considering the full breadth of LLTs (Lorentz rotations and boosts), $\udt{\Lambda}{a}{b}$, we can completely represent the spin connection as $\udt{\omega}{a}{b\mu}=\udt{\Lambda}{a}{c}\partial_{\mu}\dut{\Lambda}{b}{c}$ \cite{Krssak:2015oua}. Hence, for any metric tensor spacetime, Eq.~(\ref{metric_ten_def}) has an infinite number of solutions for the tetrad, it is the spin connection that counter-balances this and renders a covariant formulation of TG.

The Riemann tensor measures curvature, the so-called torsion tensor measures teleparallel torsion through \cite{Krssak:2018ywd}
\begin{equation}
    \udt{T}{\sigma}{\mu\nu} := 2\Gamma^{\sigma}_{[\mu\nu]}\,,
\end{equation}
where square brackets represent the usual anti-symmetric operator. The torsion tensor is a representation of the field strength of gravitation in TG \cite{Aldrovandi:2013wha}, and it transforms covariantly under both diffeomorphisms and LLTs. In order to better relate TG with Levi-Civita based theories of gravity, we necessitate the use of the contorsion tensor which is defined as the difference between the Levi-Civita and Weitzenb\"{o}ck connections, given as \cite{RevModPhys.48.393}
\begin{equation}
    \udt{K}{\sigma}{\mu\nu} := \Gamma^{\sigma}_{\mu\nu} - \mathring{\Gamma}^{\sigma}_{\mu\nu} = \frac{1}{2}\left(\dudt{T}{\mu}{\sigma}{\nu} + \dudt{T}{\nu}{\sigma}{\mu} - \udt{T}{\sigma}{\mu\nu}\right)\,,
\end{equation}
which can be represented entirely in terms of torsion tensors. Finally, the superpotential is also a useful quantity, defined as \cite{Krssak:2018ywd}
\begin{equation}
    \dut{S}{a}{\mu\nu} := \udt{K}{\mu\nu}{a} - \dut{e}{a}{\nu} \udt{T}{\alpha \mu}{\alpha} + \dut{e}{a}{\mu} \udt{T}{\alpha\nu}{\alpha}\,.
\end{equation}
The superpotential has shown to have a potential relationship with the energy-momentum tensor for gravitation \cite{Aldrovandi:2003pa,Koivisto:2019jra}. The Einstein-Hilbert Lagrangian (Ricci scalar) is the result of contractions of the Riemann tensor, TEGR is similarly the result of constrctions of the torsion tensor with its superpotential \cite{Hayashi:1979qx}
\begin{equation}
    T := \dut{S}{a}{\mu\nu}\udt{T}{a}{\mu\nu}\,,
\end{equation}
which is the result of calculations made entirely with the Weitzenb\"{o}ck connection. The Riemann tensor calculated on the Weitzenb\"{o}ck connection vanishes, which means that its Ricci scalar similarly vanishes. However, through the contorsion tensor, this can be related to the regular Ricci scalar calculated with the Levi-Civita connection, $\mathring{R}$, which results in \cite{Bahamonde:2016cul,Hayashi:1979qx,RevModPhys.48.393}
\begin{equation}
    R = \mathring{R} + T - \frac{2}{e}\partial_{\mu}\left(e\udut{T}{\sigma}{\sigma}{\mu}\right) = 0\,.
\end{equation}
This directly leads to the equivalency of the Ricci and torsion scalars (up to a total divergence term)
\begin{equation}
    \mathring{R} = -T + \frac{2}{e}\partial_{\mu}\left(e\udut{T}{\sigma}{\sigma}{\mu}\right) = -T + 2\mathring{\nabla}_{\mu}\left(\udut{T}{\sigma}{\sigma}{\mu}\right)\,,
\end{equation}
where $e=\text{det}\left(\udt{e}{a}{\mu}\right)=\sqrt{-g}$, and $B=2\mathring{\nabla}_{\mu}\left(\udut{T}{\sigma}{\sigma}{\mu}\right)$ is a total divergence term. The appearance of a boundary term makes GR and TEGR dynamically identical while retaining differences in their actions. This means that the TEGR action can be written as \cite{Aldrovandi:2013wha}
\begin{equation}
    S_{\text{TEGR}} = -\frac{1}{2\kappa^2}\int d^4 x\, eT + \int d^4 x\,e\mathcal{L}_{m}\,,
\end{equation}
where $\kappa^2=8\pi G$, and $\mathcal{L}_m$ is the Lagrangian for matter. GR and TEGR are dynamically equivalent, but the divergence term in their difference plays a crucial role in the potential modifications to their actions. Lovelock's theorem \cite{Lovelock:1971yv} limits Levi-Civita based theories of gravity to being second-order in terms of derivatives only for Lagrangians that are linear in their appearance of the Ricci scalar. TG is completely different in that it weakens this theorem and allows for generally second-order theories to exist \cite{Gonzalez:2015sha,Bahamonde:2019shr}. Using the same rationale as $f(\mathring{R})$ gravity \cite{Sotiriou:2008rp,Capozziello:2011et,Capozziello:2019cav}, the TEGR Lagrangian can be generalized to an arbitrary function of the torsion scalar, namely $f(T)$ gravity \cite{Ferraro:2006jd,Ferraro:2008ey,Bengochea:2008gz,Linder:2010py,Chen:2010va}. This is an example of a generally second-order formulation of TG where the resulting field equations will be second-order in tetrad derivatives irrespective of the form of the Lagrangian function.

More generally, using the same reasoning as $f(\mathring{R})$ gravity, we can generalize the TEGR Lagrangian to an arbitrary function of both the torsion scalar as well as the associated boundary quantity that relates TEGR and GR, i.e. $f(T,B)$. In this way, the torsion scalar embodies the second-order contributions to the field equations while the boundary term incorporates the fourth-order contributions \cite{Bahamonde:2015zma}. For this reason, $f(T,B)$ gravity emerges as a generalization of $f(\mathring{R})$ gravity where equivalency results for the particular choice of $f(\mathring{R}) = f(-T+B)$.

Thus, $f(T,B)$ gravity acts as a novel approach to gravitation in which the distinct derivative contributions acts independently of each other. By taking a variation of this modified Lagrangian, the following field equations are found \cite{Bahamonde:2015zma,Farrugia:2018gyz}
\begin{align}
  & 2\delta_{\nu}^{\lambda}\Box f_{B}-2\nabla^{\lambda}\nabla_{\nu}f_{B}+
  e B f_{B}\delta_{\nu}^{\lambda} + 
  4\Big[(\partial_{\mu}f_{B})+(\partial_{\mu}f_{T})\Big]S_{\nu}{}^{\mu\lambda}  \nonumber\\
  & +4e^{-1}e^{a}{}_{\nu}\partial_{\mu}(e S_{a}{}^{\mu\lambda})f_{T} - 
  4  f_{T}T^{\sigma}{}_{\mu \nu}S_{\sigma}{}^{\lambda\mu} - 
   f \delta_{\nu}^{\lambda} = 2\kappa^2 \dut{\Theta}{\nu}{\lambda} \,.
  \label{FieldEq}
\end{align}
where subscripts denote derivatives, and $\Theta_{\nu\lambda}=\udt{e}{a}{\nu}\dut{\Theta}{a}{\lambda}$ is the regular energy-momentum tensor for matter. As in standard gravity, we can then probe the theory by considering different spacetimes. In the following sections, we investigate the effect of this modified theory of gravity for spherically symmetric spacetimes.

\section{Spherical symmetry in $f(T,B)$ gravity}\label{sec:sph}
This section will be devoted to derive the field equations in spherical symmetry respecting LLT invariance and then to find different perturbed solutions around Schwarzschild for different power-law forms of $f(T,B)$ gravity.

\subsection{Field equations and basic ingredients}
Let us assume a spherically symmetric spacetime whose metric is
\begin{equation}\label{metric_ansatz}
    ds^2=\mathcal{A}(r)dt^2-\mathcal{B}(r)dr^2-r^2d\Omega^2\,,
\end{equation}
where $\mathcal{A}(r)$ and $\mathcal{B}(r)$ are positive functions, which is reproduced by the off-diagonal tetrad
\begin{equation}\label{tetrad_ansatz}
    e^a{}_\mu=\left(
\begin{array}{cccc}
 \sqrt{\mathcal{A}} & 0 & 0 & 0 \\
 0 & \sqrt{\mathcal{B}} \sin (\theta ) \cos (\phi ) & r \cos (\theta ) \cos (\phi ) & -r \sin (\theta ) \sin (\phi ) \\
 0 & \sqrt{\mathcal{B}} \sin (\theta ) \sin (\phi ) & r \cos (\theta ) \sin (\phi ) & r \sin (\theta ) \cos (\phi ) \\
 0 & \sqrt{\mathcal{B}} \cos (\theta ) & -r \sin (\theta ) & 0 \\
\end{array}
\right)\,.
\end{equation}
The torsion scalar and the boundary term for this tetrad behave as
\begin{align}
    T&=-\frac{2 \left(\sqrt{\mathcal{B}}-1\right) \left(r \mathcal{A}'-\mathcal{A} \sqrt{\mathcal{B}}+\mathcal{A}\right)}{r^2 \mathcal{A} \mathcal{B}}\,,\\
    B&=\frac{-r^2 \mathcal{B} \mathcal{A}'^2+r \mathcal{A} \left(-r \mathcal{A}' \mathcal{B}'-4 \mathcal{B}^{3/2} \mathcal{A}'+2 \mathcal{B} \left(r \mathcal{A}''+4 \mathcal{A}'\right)\right)-4 \mathcal{A}^2 \left(r \mathcal{B}'+2 \mathcal{B}^{3/2}-2 \mathcal{B}\right)}{2 r^2 \mathcal{A}^2 \mathcal{B}^2}\,.
\end{align}
It is important to remark that the tetrad \eqref{tetrad_ansatz} is a good tetrad in the sense that is compatible with a vanishing spin connection and also both the scalar torsion and the boundary term are zero for the Minkowski case ($\mathcal{A}=1$ and $\mathcal{B}=1$). For this tetrad, the $f(T,B)$ field equations \eqref{FieldEq} yield
\begin{eqnarray}
\frac{1}{2} \kappa ^2 \rho &=&\frac{1}{4} f+\frac{ r \mathcal{B} (\sqrt{\mathcal{B}}-1) \mathcal{A}'+\mathcal{A} (r \mathcal{B}'+2 \mathcal{B}^{3/2}-2 \mathcal{B})}{2 r^2 \mathcal{A} \mathcal{B}^2}f_T-\frac{r \mathcal{B}' f'_B-4 \mathcal{B}^{3/2} \left(f'_B+f'_T\right)+4 \mathcal{B} f'_T}{4 r \mathcal{B}^2}+\frac{f''_B}{2 \mathcal{B}}\,,\nonumber\\
&&+\frac{r^2 \mathcal{B} \mathcal{A}'^2+r \mathcal{A} \Big[r \mathcal{A}' \mathcal{B}'+4 \mathcal{B}^{3/2} \mathcal{A}'-2 \mathcal{B} (r \mathcal{A}''+4 \mathcal{A}')\Big]+4 \mathcal{A}^2 (r \mathcal{B}'+2 \mathcal{B}^{3/2}-2 \mathcal{B})}{8 r^2 \mathcal{A}^2 \mathcal{B}^2}f_B\,,\label{Eq1}\\
\frac{1}{2} \kappa ^2 p_r&=&-\frac{1}{4} f+\frac{ -r^2 \mathcal{B} \mathcal{A}'^2+r \mathcal{A} \left[-r \mathcal{A}' \mathcal{B}'-4 \mathcal{B}^{3/2} \mathcal{A}'+2 \mathcal{B} (r \mathcal{A}''+4 \mathcal{A}')\right]-4 \mathcal{A}^2 \left(r \mathcal{B}'+2 \mathcal{B}^{3/2}-2 \mathcal{B}\right)}{8 r^2 \mathcal{A}^2 \mathcal{B}^2}f_B-\frac{r \mathcal{A}'+4 \mathcal{A} }{4 r \mathcal{A} \mathcal{B}}f'_B\nonumber\\
&&-\frac{ r (\sqrt{\mathcal{B}}-2) \mathcal{A}'+2 \mathcal{A} (\sqrt{\mathcal{B}}-1)}{2 r^2 \mathcal{A} \mathcal{B}}f_T\,,\label{Eq2}\\
\frac{1}{2} \kappa ^2 p_l&=&-\frac{1}{4} f+\frac{r \mathcal{A}'-2 \mathcal{A} (\sqrt{\mathcal{B}}-1)}{4 r \mathcal{A} \mathcal{B}}f'_T+\frac{r \mathcal{B}'-2 \mathcal{B}^{3/2} }{4 r \mathcal{B}^2}f'_B-\frac{f''_B}{2 \mathcal{B}}\nonumber\\
&&+\frac{-r^2 \mathcal{B} \mathcal{A}'^2+r \mathcal{A} \left[-r \mathcal{A}' \mathcal{B}'-4 \mathcal{B}^{3/2} \mathcal{A}'+2 \mathcal{B} (r \mathcal{A}''+3 \mathcal{A}')\right]+\mathcal{A}^2 (-2 r \mathcal{B}'-8 \mathcal{B}^{3/2}+4 \mathcal{B}^2+4 \mathcal{B})}{8 r^2 \mathcal{A}^2 \mathcal{B}^2}f_T\nonumber\\
&&+\frac{-r^2 \mathcal{B} \mathcal{A}'^2+r \mathcal{A} \left[-r \mathcal{A}' \mathcal{B}'-4 \mathcal{B}^{3/2} \mathcal{A}'+2 \mathcal{B} (r \mathcal{A}''+4 \mathcal{A}')\right]-4 \mathcal{A}^2 (r \mathcal{B}'+2 \mathcal{B}^{3/2}-2 \mathcal{B})}{8 r^2 \mathcal{A}^2 \mathcal{B}^2}f_B \ \,,\label{Eq3}
\end{eqnarray}
where primes denote differentiation with respect to the radial coordinate, then, $f'_T=f_{TT}T'+f_{TB}B'$ and $f'_B=f_{BB}B'+f_{TB}T'$. Clearly, if we set $f=f(T)$, we recover the field equations studied in \cite{Bahamonde:2019zea} and if $f=f(-T+B)=f(\lc{R})$ we recover the spherically symmetric equations in $f(\lc{R})$ gravity reported in~Refs.~\cite{Capozziello:2007wc,Bahamonde:2019swy}. In the above equations, we have assumed that the matter is described by an anisotropic fluid with an energy density $\rho$ and radial and lateral pressures $p_r$ and $p_l$ respectively. This system of three differential equations only have two that are independent, and similarly as in $f(T)$ gravity~\cite{Golovnev:2020las}, it can be noticed that the Bianchi identities are satisfied too since the antisymmetric equations are also satisfied. 

Let us remark here that the use of the tetrad~\eqref{tetrad_ansatz} with a zero spin connection (as we did here) is equivalent as having a diagonal tetrad with a non-zero spin connection as it was done in~\cite{Hohmann:2019nat,Krssak:2015oua}. These approaches are equivalent since they are related via a Lorentz transformations. The important point is to use the correct tetrad which is compatible with a zero spin connection that also satisfies the symmetries of the underlying metric~\cite{Hohmann:2019nat}. It is also important to emphasise that it does not matter which tetrad-spin connection pair one uses, the field equations and then the Solar System constraints will be the same. This is again due to the fact that the pair tetrad-spin connection can be always related via a Lorentz transformations but with the important point that they must respect the symmetries of the problem~\cite{Hohmann:2019nat}.

\subsection{Perturbed spherically symmetric solutions\label{pert_sols}}
This section will be devoted to find perturbed spherically symmetric solutions for different types of $f(T,B)$ theories of gravity. We will use the same approach as in \cite{Bahamonde:2019zea} where it was assumed that the background is described by the Schwarzschild geometry and the perturbed coefficients are first order corrections to this spacetime, namely,
\begin{align}
    \mathcal{A}(r)&=1-\frac{2M}{r}+\epsilon\, a(r)\,,\label{AA}\\
    \mathcal{B}(r)&=\Big(1-\frac{2M}{r}\Big)^{-1}+\epsilon \,b(r)\,.\label{BB}
\end{align}
Here $\epsilon\ll 1$ is a small tracking parameter that is used to make series expansions in a coherent way. This is not to be confused with a model parameter which does have a physical impact, $\epsilon$ is simply used to keep track of leading order terms in a straightforward way. Hereafter, we will assume that there is no additional matter, i.e., we will find perturbed solutions in vacuum $\rho=p_r=p_l=0$. To include different power-law forms of the Lagrangian, let us assume the following combination of power-law terms
\begin{equation}\label{model_choice}
   f(T,B)= T+\frac{1}{2} \epsilon  \left(\alpha  T^q+\beta  B^m+\gamma  B^s T^w+\zeta  (\xi  T+\chi B  )^u\right)\,,
\end{equation}
where $\alpha,\beta,\gamma,\zeta,q,m,s,w$ and $u$ are constants. In order to get the perturbed equations for this model, we need to replace the above form of $f$ and the metric coefficients \eqref{AA} and \eqref{BB} into the field equations \eqref{Eq1}--\eqref{Eq3} and then expand them up to first order in $\epsilon$. The corresponding expanded equations are cumbersome so we present them in the Appendix (see Eqs.~\eqref{eqapp1} and \eqref{eqapp2}).  It is only for some set of parameters that we can find solutions for these equations. We will now split the study for different cases depending on the constants in which we can solve them. We will find six different solutions that can be categorised into two cases: 
\begin{enumerate}
    \item Case 1: ($\zeta=0, \, q=m=2$)
    \begin{enumerate}
        \item $f(T,B)=T+\frac{1}{2} \epsilon  \left(\alpha  T^2+\beta  B^2+\gamma  B^s T^{1-s}\right)$, meaning $w=-s+1$ 
        \item $f(T,B)=T+\frac{1}{2} \epsilon  \left(\alpha  T^2+\beta  B^2+ \gamma  B  T\right)$, meaning $w=s=1$
         \item $f(T,B)=T+\frac{1}{2} \epsilon  \left(\alpha  T^2+\beta  B^2+\gamma B^2   T\right)$, meaning $w=1,\, s=2$
              \item $f(T,B)=T+\frac{1}{2} \epsilon  \left(\alpha  T^2+\beta  B^2+ \gamma  B  T^2\right)$, meaning $w=2,\, s=1$
    \end{enumerate}
     \item Case 2: ($\alpha=\beta=\gamma=0$)
    \begin{enumerate}
    \item  $f(T,B)= T+\frac{1}{2} \epsilon \, \zeta  (\xi  T+\chi B  )^3$, meaning $u=3$
    \item  $f(T,B)= T+\frac{1}{2} \epsilon\,  \zeta  (\xi  T+\chi B  )^4$, meaning $u=4$
    \end{enumerate}
\end{enumerate}
\noindent It should be remark that for all the cases that we will study, the Minkowski background case, $M=0$, only gives trivial solutions, i.e., that the perturbed terms do not contribute up to first order in $\epsilon$.

\subsubsection{Case 1: $\zeta=0,\, q=m= 2$}
As a first case, we take the constants $\zeta=0,\, q=m= 2$, which gives the following form of the Lagrangian
\begin{equation}
    f(T,B)=T+\frac{1}{2} \epsilon \, \left(\alpha  T^2+\beta  B^2+\gamma B^sT^w\right)\,,
\end{equation}
which is a generalisation of the squared $f(T)$ model studied in~\cite{Bahamonde:2019zea}. For this model, we can find different set of solutions for the system~\eqref{eqapp1}-\eqref{eqapp2} depending on $s$ and $w$. One can write them as follows,
\begin{eqnarray}
\mathcal{A}(r)&=&\mu^2+\epsilon\Big[C_2-\frac{1}{r^2(\mu^2-1)^2}\Big(3 \beta  \mu^7-\frac{1}{2} (\alpha +13 \beta ) \mu^6-4 \beta  \mu^5+\frac{1}{2} \mu^4 (15 \alpha +43 \beta +2 C_1r)-\frac{2}{3}\mu^3 (32 \alpha +35 \beta ) \nonumber\\
&&-\frac{1}{2} \mu^2 (31 \alpha +51 \beta +4 C_1r)+4 \beta  \mu+\frac{1}{2} (17 \alpha +21 \beta +2 C_1r)-\frac{\beta }{\mu}+2 (\alpha +\beta ) \left(3 \mu ^2-1\right) \log\mu \Big)\Big]\nonumber\\
&&+\epsilon\, \tilde{a}_{\gamma}(r)\,,\label{solA}\\
\mathcal{B}(r)&=&\mu^{-2}+\epsilon\Big[\frac{C_1}{r \mu^4}+\frac{C_2 \left(\mu^2-1\right)}{\mu^4}+\frac{1}{r^2(\mu^2-1)}\Big(\frac{1}{2} (25 \alpha +37 \beta )-4 (\alpha +2 \beta ) \mu-\frac{2 (16 \alpha +13 \beta )}{3 \mu}-\frac{2 (\alpha +3 \beta )}{\mu^2}\nonumber\\
&&+\frac{4 (\alpha +\beta )}{\mu^3}+\frac{-21 \alpha -25 \beta }{2 \mu^4}+\frac{2 \beta }{\mu^5}+\frac{2 (\alpha +\beta ) }{\mu^4}\log\mu\Big)\Big]+\epsilon\, \tilde{b}_{\gamma}(r)\,,\label{solB}
\end{eqnarray}
where $\mu^2=1-2M/r$, $C_1$ and $C_2$ are integration constants, and the value of $\tilde{a}_{\gamma}(r)$ and $\tilde{b}_{\gamma}(r)$ depend on $\gamma$ (or $s$ and $w$). We can find four different sets of solutions depending on the choice of $w$ and $s$, namely, (i) Case 1a: $w=-s+1$; (ii) Case 1b: $w=s=1$, (iii) Case 1c: $w=1, s=2$, (iv) Case 1d: $w=2, s=1$. To get the value of the constants $C_1$ and $C_2$, we need to expand the above metric coefficients up to $1/r$. For the first case, $w=-s+1$, the resulting solution does not depend on the parameter $\gamma$, so that, there is no contribution of the term related to $\gamma  B^s T^{1-s}$ up to first order in the perturbation. Expanding up to $1/r$, for each case we find
\begin{eqnarray}
 a(r)&\sim& \Big(C_2+\frac{16 (\alpha +\beta )}{3 M^2}-C_{2,\gamma}\Big)-\Big(C_1+\frac{16 (\alpha +\beta )}{M}-C_{1,\gamma}\Big)\frac{1}{r}+\mathcal{O}\Big(\frac{1}{r^2}\Big) \,,\\
 b(r)&\sim& \Big(C_1-2 C_2 M+\frac{16 (\alpha +\beta )}{3 M}+C_{1,\gamma}+2 C_{2,\gamma} M\Big)\frac{1}{r}+\mathcal{O}\Big(\frac{1}{r^2}\Big)\,,
\end{eqnarray}
where $C_{1,\gamma}$ and $C_{2,\gamma}$ are constants depending on the model. To get the correct limit up to $1/r$ order ($a_1=0, b_1=0$), we must impose
\begin{equation}
    C_1=-\frac{16\alpha +\beta}{M}+C_{1,\gamma}\,, \quad C_2=-\frac{16 \alpha +\beta}{3 M^2}+C_{2,\gamma}\,.
\end{equation}
For each case, we would have different constants $C_1$ and $C_2$ since the constants $C_{1,\gamma}$ and $C_{2,\gamma}$ depend on the model. By replacing these values for $C_1$ and $C_2$ into the solution \eqref{solA} and \eqref{solB}, we get the following perturbed solutions,
\begin{eqnarray}
\mathcal{A}(r)&=&\mu^2-\frac{1}{(\mu^2-1)^2r^2}\Big[3 \beta  \mu^7-\frac{1}{2} (\alpha +13 \beta ) \mu^6-4 \beta  \mu^5+\frac{1}{2} (15 \alpha +43 \beta ) \mu^4-\frac{2}{3} (32 \alpha +35 \beta ) \mu^3+\frac{1}{2} (33 \alpha +13 \beta ) \mu^2\nonumber\\
&&+4 \beta  \mu-\frac{1}{6} (13 \alpha +\beta )-\frac{\beta }{\mu}-2 (\alpha +\beta ) \left(1-3 \mu^2\right)\log\mu\Big]+\epsilon\, \tilde{a}_{\gamma}(r)\,,\\
\mathcal{B}(r)&=&\mu^{-2}+\frac{\epsilon}{r^2(\mu^2-1)}\Big[\frac{1}{2} (25 \alpha +37 \beta )-4 (\alpha +2 \beta ) \mu-\frac{2 (16 \alpha +13 \beta )}{3 \mu}-\frac{2 (\alpha +3 \beta )}{\mu^2}+\frac{4 (\alpha +\beta )}{\mu^3}\nonumber\\
&&+\frac{\alpha -11 \beta }{6 \mu^4}+\frac{2 \beta }{\mu^5}+\frac{2 (\alpha +\beta ) \log (\mu)}{\mu^4}\Big]+\epsilon \,\tilde{b}_{\gamma}(r)\,,
\end{eqnarray}
where for each case, we find that the functions $\tilde{a}_{\gamma}(r)$ and $\tilde{b}_{\gamma}(r)$ become
\begin{itemize}
    \item Case 1a: $w=-s+1$ yielding $f(T,B)=T+\frac{1}{2} \epsilon  \left(\gamma  B^s T^{1-s}+\beta  B^2+\alpha  T^2\right)$:  \begin{equation}
        \tilde{a}_{\gamma}(r)=\tilde{b}_{\gamma}(r)=0\,\,,\quad C_{1,\gamma}=C_{2,\gamma}=0
    \end{equation}
    \item Case 1b: $w=s=1$ yielding $f(T,B)=T+\frac{1}{2} \epsilon  \left(\beta  B^2+B \gamma  T+\alpha  T^2\right)$: 
    \begin{eqnarray}
\tilde{a}_{\gamma}(r)&=&\frac{\gamma}{r^2 \left(\mu^2-1\right)^2}\Big[-\frac{3}{2}\mu^7+\frac{7 \mu^6}{2}+2 \mu^5-\frac{29 \mu^4}{2}+\frac{67 \mu^3}{3}-\frac{23 \mu^2}{2}-2 \mu+\frac{1}{2 \mu}+\frac{7}{6}+2\left(1-3 \mu^2\right) \log (\mu)\Big]\,,\nonumber \\ && \\
\tilde{b}_{\gamma}(r)&=&\frac{\gamma}{r^2(\mu^2-1)}\Big[-6 \mu-\frac{29}{3 \mu}-\frac{4}{\mu^2}+\frac{4}{\mu^3}-\frac{5}{6 \mu^4}+\frac{1}{\mu^5}+\frac{31}{2}+\frac{2 \log (\mu)}{\mu^4}\Big]\,,\\
C_{1,\gamma}&=&-\frac{16 \gamma }{M}\,,\quad C_{2,\gamma}=-\frac{16 \gamma }{3 M^2}\,.
\end{eqnarray}
 \item Case 1c: $w=1, s=2$ yielding $f(T,B)=T+\frac{1}{2} \epsilon  \left(\beta  B^2+B^2 \gamma  T+\alpha  T^2\right)$
   \begin{eqnarray}
\tilde{a}_{\gamma}(r)&=&\frac{\gamma}{r^4  \left(\mu^2-1\right)^4}\Big[6 \mu^{12}-\frac{224 \mu^{11}}{9}+13 \mu^{10}+\frac{752 \mu^9}{9}-\frac{397 \mu^8}{3}-\frac{1744 \mu^7}{35}+\frac{778 \mu^6}{3}-\frac{2272 \mu^5}{15}\nonumber\\
&&-154 \mu^4+\frac{1088 \mu^3}{3}-\frac{6917 \mu^2}{45}-80 \mu-\frac{2}{\mu^2}+\frac{16}{\mu}+\frac{2417}{315}+24\left(1-7 \mu^2\right) \log (\mu)\Big]\,,\\
\tilde{b}_{\gamma}(r)&=&\frac{\gamma}{r^4 \left(\mu^2-1\right)^3}\Big[30 \mu^6-\frac{1120 \mu^5}{9}+115 \mu^4+\frac{1376 \mu^3}{7}-\frac{1286 \mu^2}{3}+\frac{576 \mu}{5}-\frac{832}{3 \mu}+\frac{42}{\mu^2}+\frac{32}{\mu^3}-\frac{10813}{315 \mu^4}\nonumber\\
&&+\frac{32}{\mu^5}-\frac{6}{\mu^6}+308+\frac{24 \log (\mu)}{\mu^4}\Big]\,,\\
C_{1,\gamma}&=&-\frac{1024 \gamma }{45 M^3}\,,\quad C_{2,\gamma}=-\frac{1024 \gamma }{105 M^4}\,.
\end{eqnarray}
 \item Case 1d: $w=2, s=1$ yielding $f(T,B)=T+\frac{1}{2} \epsilon  \left(\beta  B^2+B \gamma  T^2+\alpha  T^2\right)$:
    \begin{eqnarray}
\tilde{a}_{\gamma}(r)&=&\frac{\gamma}{r^4 \left(\mu^2-1\right)^4}\Big[3 \mu^{12}-\frac{116 \mu^{11}}{9}+8 \mu^{10}+\frac{392 \mu^9}{9}-\frac{238 \mu^8}{3}-\frac{484 \mu^7}{35}+\frac{475 \mu^6}{3}-\frac{2032 \mu^5}{15}-75 \mu^4+\frac{956 \mu^3}{3}\nonumber\\
&&-\frac{7952 \mu^2}{45}-56 \mu-\frac{1}{\mu^2}+\frac{12}{\mu}+\frac{2102}{315}+24\left(1-7 \mu^2\right) \log (\mu)\Big]\,,\\
\tilde{b}_{\gamma}(r)&=&\frac{\gamma}{r^4 \left(\mu^2-1\right)^4}\Big[21 \mu^8-\frac{832 \mu^7}{9}+80 \mu^6+\frac{13672 \mu^5}{63}-\frac{1334 \mu^4}{3}+\frac{792 \mu^3}{35}+\frac{1703 \mu^2}{3}-\frac{6128 \mu}{15}\nonumber\\
&&+\frac{880}{3 \mu}-\frac{28768}{315 \mu^2}-\frac{8}{\mu^3}+\frac{9238}{315 \mu^4}-\frac{24}{\mu^5}+\frac{3}{\mu^6}-165+24\left(\frac{1}{\mu^2}-\frac{1}{\mu^4}\right) \log (\mu)\Big]\,,\\
C_{1,\gamma}&=&-\frac{1024 \gamma }{45 M^3}\,,\quad C_{2,\gamma}=-\frac{1024 \gamma }{105 M^4}\,.
\end{eqnarray}
\end{itemize}
It is worth mentioning that for the Case 1b, if one takes $\gamma=-2\beta$ and $\alpha=\beta$, the model becomes $f(T,B)=T+(1/2)\epsilon \,\beta \lc{R}{}^2$ which is equivalent as having GR plus a squared power-law $f(\lc{R})$. Obviously, for this case we find that there are no $\epsilon$ corrections (up to first order) for a Schwarzschild background, i.e, $\mathcal{A}(r)=1/\mathcal{B}(r)=\mu^2=1-2M/r$. Let us remark here that all the above solutions are asymptotically flat, i.e, when $r\rightarrow \infty$, one finds $\mathcal{A}(r)\rightarrow 0$ and $\mathcal{B}(r)\rightarrow 0$.
\subsubsection{Case 2: $\alpha=\beta=\gamma=0$}
As a second case, we will assume that the constants  $\alpha=\beta=\gamma=0$ yielding the following form of $f$ ($\xi\neq\chi$):
\begin{equation}
    f(T,B)=T+\frac{1}{2}\epsilon\, \zeta  (\xi  T+\chi B  )^u\,, \quad u\in \mathcal{N}\quad \textrm{and}\quad u\neq1\,.
\end{equation}
The specific case $u=2$ is equivalent to the Case 1b studied before with $\gamma=2\sqrt{\alpha\beta}, \alpha=\chi^2, \beta=\xi^2$ and $\zeta=1$. For all other values of $u$, it is possible to solve the system for $a$ and $b$.  As interesting examples, in the Appendix (see Sec.~\ref{appendix33}), we show the solutions for the $u=3$ and $u=4$ cases.

\section{Solar System Tests}\label{sec:solar}
GR theory is very successful in predicting the behavior of the gravitational phenomena in the Solar System, so every kind of generalization of GR must be  consistent in the Solar System as a minimal requirement to be physically viable more generally. There are many experiments that could be used to test gravity theories in a relatively high accurate level, including those in the Solar system, such as perihelion shift, deflection of light, Cassini experiment, Shapiro delay and the gravitational redshift. Here, we apply these tests to $f(T,B)$ perturbed solutions presented in Sec.~\ref{sec:sph}.

We can write down the worldline $q(\tau)$ of a test particle as
\begin{equation}
2\mathcal{L}=g_{\mu\nu}\dot{q}^{\mu}\dot{q}^{\nu}=\mathcal{A}\,\dot{t}^2-\mathcal{B}\,\dot{r}^2-r^2\dot{\theta}^2-r^2\sin^2{\theta}\dot{\phi}^2\,,\label{EL}
\end{equation}
where $q^{\mu}(\tau)=(t(\tau), r(\tau), \theta(\tau), \phi(\tau))$ and dots denote derivative with respect to the affine parameter $\tau$. We restrict ourselves to motion in the equatorial plane and set $\theta=\pi/2$, and then we find that the conserved quantities, energy $k$ and momentum $h$ become, respectively
\begin{eqnarray}\label{kh}
k &=& \frac{\partial \mathcal{L}}{\partial \dot t} =\mathcal{A}\dot{t}= \left(1-\frac{2 M}{r}+ \epsilon\, a(r)\right)\dot{t}\,,\\
	h &=& -\frac{\partial \mathcal{L}}{\partial \dot \phi} = r^2 \dot \phi\,.
\end{eqnarray}
Using these quantities in Eq.~\eqref{EL}, we find
\begin{align}\label{17}
\dot{r}^2=\mathcal{B}^{-1}\Big(\frac{k^2}{\mathcal{A}}-\frac{ h^2}{r^2}-\sigma\Big)\,.
\end{align}
Here, $\sigma=1$ for massive particles and $\sigma=0$ for massless particles.  We can further rewrite the above equation as follows
\begin{align}
	\dot r^2 + 2V(r)=0\,,\label{potentialEq}
\end{align}
where we have defined the potential as
\begin{align}\label{eq:pot1}
	V(r) &=-\frac{1}{2}\mathcal{B}^{-1}\Big(\frac{k^2}{\mathcal{A}}-\frac{ h^2}{r^2}-\sigma\Big)\,.
\end{align}
By replacing the metric functions and expanding up to first order in $\epsilon$, we get that the potential becomes
\begin{align}\label{eq:pot}
	V(r) &= - \frac{1}{2} k^2 + \frac{1}{2} \left(1-\frac{2M}{r}\right) \left( \frac{h^2}{r^2} + \sigma \right) \nonumber\\
&+ \frac{\epsilon}{2} \left[ k^2 \left( \frac{a(r)}{1-\frac{2M}{r}} + b(r)\left(1-\frac{2M}{r}\right)\right) - b(r)\left( \sigma + \frac{h^2}{r^2} \right)\left(1-\frac{2M}{r}\right)^2\right]\,.
\end{align}
Having this potential, we can now analyse different observable that one can constrain using Solar System data.

\subsection{Photon sphere and perihelion shift}\label{geodesics}
This section will be devoted to studying the photon sphere and the perihelion shift for each perturbed solution found in the previous section. For circular photon orbits ($\sigma=0$) we must have that the potential and its derivatives vanish, i.e., $V=V'=0$. By expanding the radial (circular) coordinate $r_c=r_0+\epsilon\, r_1$, energy $k=k_0+\epsilon\, k_1$ and angular momentum $h=h_0+\epsilon\, h_1$ and then by solving order by order into the conditions $V=V'=0$, we obtain that, the zeroth-order gives the standard GR term
\begin{align}
	r_0 = 3M,\quad h_{0\pm} = \pm 3 \sqrt{3} k_0 M\,,
\end{align} 
and the first order correction are
\begin{align}
    r_1&=\frac{4 \left(\sqrt{3}-9\right) \alpha -2 \left(\sqrt{3}+15\right) \beta }{9 \left(\sqrt{3}+3\right) M}+\frac{3 \log (3) (\alpha +\beta )}{4 M}+\tilde{r}_1\approx \frac{0.141338 \alpha }{M}+\frac{0.038204 \beta }{M}+\tilde{r}_1\,, \\ h_{1\pm}&=\pm\frac{2 \left(3 \sqrt{3}-5\right) k_0 (\alpha +\beta )}{\left(\sqrt{3}+3\right) M}\pm\frac{9 \left(\sqrt{3}+1\right) k_1 M}{\sqrt{3}+3}+\tilde{h}_{1\pm}\approx\pm\Big(\frac{0.0829 k_0 (\alpha +\beta )}{M}+5.196 k_1 M\Big)+\tilde{h}_{1\pm}\,,
\end{align}
where the terms $\tilde{r}_1$ and $\tilde{h}_{1\pm}$ depend on the model. Their values are displayed in the Appendix~\ref{photonsphere}. The photon sphere defines the edge of the shadow of the black hole. One can notices that  depending on the parameters, the photon sphere could be larger or smaller than the standard result in GR. Explicitly, the shadow of the black hole predicted by power-law $f(T,B)$ will be enlarged for each solution if
\begin{eqnarray}
\textrm{ \textbf{Case 1a:}}&&\ \alpha + 0.270 \beta\gtrsim 0  \,, \label{sh_case1a}\\  \textrm{ \textbf{Case 1b:}}&&\ \ \alpha + 0.270 \beta+0.635 \gamma\gtrsim 0\,,\label{sh_case1b}\\
\textrm{ \textbf{Case 1c:}}&&\ \ \alpha + 0.270 \beta-\frac{0.04623}{M^2} \gamma\gtrsim 0\,,\label{sh_case1c}\\
\textrm{\textbf{ Case 1d:}}&&\ \ \alpha + 0.270 \beta-\frac{0.0713}{M^2} \gamma\gtrsim 0\,,\label{sh_case1d}\\
 \textrm{ \textbf{Case 2a:}}&&\ \ -1.363\times 10^{-2} \zeta  (\xi +0.219 \chi )\gtrsim 0\,,\label{sh_case2a}\\
 \textrm{\textbf{Case 2b:}}&&\ \ 
1.213\times 10^{-3} \zeta  (\xi+1.962\times 10^{-1}\chi) (\xi +\chi )^3\gtrsim 0\,,\label{sh_case2b}
\end{eqnarray}
and the shadow will be reduced if we take the opposite inequalities ($\lesssim$). It is also interesting to note that it is also possible to find specific constants such as the shadow of the black holes would be the same as GR. This case would happen when all the above quantities are $\approx 0$ instead of $\gtrsim 0$. Let us know study what happens if we just keep one of the constants for each model and we set the others to zero. The term $T^2$ is related to $\alpha$, so that one can conclude that the shadow of the black holes will be enlarged (reduced) if $\alpha>0$ ($\alpha<0$). The same happens for the $B^2$  contribution, i.e., when $\beta>0$ ($\beta<0$), the shadow will be bigger. The same situation happens for the term $\gamma B T$ (Case 1b), where $\gamma>0$ ($\gamma<0$) will enlarge (reduce) the photon sphere. On the contrary for the contributions $\gamma B T^2$ (Case 1c) and $\gamma B^2 T$ (Case 1d), one needs $\gamma<0$ ($\gamma>0$) for a larger (smaller) shadow. The Case 2 involves two parameters $\chi$ and $\xi$,  which exhibit different behavior in both subcases (i.e., 2a, 2b). In Case 2a, positive $\chi$ and $\xi$ predict smaller black hole shadow, while in the Case 2b signs of $\chi$ and $\xi$ become irrelevant and $f(T,B)$ perturbation yields larger photon sphere around the black hole for any value of $\chi$ and $\xi$. Here, we can clearly see that by increasing the value of the power $u$ (where $f(T,B)=T+ (1/2)\zeta (\chi T+\xi B)^u$), decrease the influence of the modification coming from the perturbed $f(T,B)$ gravity.

Let us now study the situation of the circular orbits for massive particles ($\sigma=1$). One of the most interesting quantities for this situation is to studying how the perihelion shift changes with respect to the quantity measured in GR. This quantity can be written as follows~\cite{Bahamonde:2019zea}
\begin{align}
    \Delta \phi =2\pi\Big(\frac{1}{K}-1\Big) =2\pi \left(\frac{h}{r_c^2\sqrt{V''(r_c)}}-1\right)\,.
\end{align}
where $K$ is the wave number, and $r_c$ is a perturbation around a circular orbit described by $r(\phi)=r_c+r_\phi(\phi)$. The above equation was computed by assuming that the ratio $r_\phi/r_c\ll 1$ is small and that the potential $V(r_c)=0$ and its derivative $V'(r_c)=0$ vanish at $r_c$. For further details about these calculations, see~\cite{Bahamonde:2019zea}. For all the perturbed solutions found in the previous section, we find that the perihelion shift is given by
\begin{eqnarray}\label{eqn:perihelionshift}
    \Delta\phi(h_{0+},h_{1+}) &=& \Delta\phi_{\rm GR}+ \epsilon\, \Delta\phi_{\epsilon} \\
    &=&6 \pi q + 27 \pi q^2 +135 \pi  q^3+ \mathcal{O}(q^4) + \epsilon\,\pi \Big(\frac{12   \beta  q}{r_c^2}+\frac{8   q^2 (\alpha +10 \beta )}{r_c^2}+\frac{  q^3 (194 \alpha +1139 \beta )}{2 r_c^2}+\gamma\Delta\phi_{\gamma}+\zeta\Delta\phi_{\zeta}\Big)\,,\nonumber \\
\end{eqnarray}
where $q=M/r_c$ and for each model we have,
    \begin{eqnarray}
\textrm{ \textbf{Case 1a:}}&&\ \ \gamma=\zeta=0\,, \label{pw_case1a}\\  \textrm{ \textbf{Case 1b:}}&&\ \ \Delta\phi_{\gamma}=\frac{44  q^2}{r_c^2}+\frac{6  q}{r_c^2}+\frac{1333  q^3}{4 r_c^2}\,,\quad \zeta=0\,,\label{pw_case1b}\\
\textrm{ \textbf{Case 1c:}}&&\ \ \Delta\phi_{\gamma}=-\frac{112  q^3}{r_c^4}\,,\quad \zeta=0\,,\label{pw_case1c}\\
\textrm{\textbf{ Case 1d:}}&&\ \ \Delta\phi_{\gamma}=-\frac{56   q^3}{r_c^4}\,,\quad \zeta=0\,,\label{pw_case1d}\\
 \textrm{ \textbf{Case 2a:}}&&\ \ \Delta\phi_{\zeta}=-\frac{168   q^3 \chi  (\xi +\chi )^2}{r_c^4}\,, \quad  \alpha=\beta=\gamma=0\,,\label{pw_case2a}\\
 \textrm{\textbf{Case 2b:}}&&\ \ \Delta\phi_{\zeta}=\frac{1056   q^5 \chi  (\xi +\chi )^3}{r_c^6}\,, \quad \alpha=\beta=\gamma=0\,.\label{pw_case2b}
\end{eqnarray}
It should be noted that the leading term for the Case 2a is $q^5$, so that its correction is very small. Since this case has a very small correction, we are only showing up to $q^5$ for all the other models. In the next sections, we will see a similar behaviour, i.e., the Case 2a only has very tiny corrections to the expansions.

Let us now consider the perihelion shift for Mercury, which has a $r_c=r_{c,\rm Mercury}\approx 5.550\cdot 10^7$ km and $M=M_{\odot}\approx 1.474\, \textrm{km}$ (Mass of the Sun in units of the Schwarzschild radius). The perihelion shift $ \Delta\phi$ is in radians over cycles, so we also need the period of Mercury around the Sun which is approximately $T_{\rm Mercury}\approx 87,97\, \textrm{days}\approx 2.410\cdot 10^{-3} \textrm{centuries}$. Using these numbers, one finds that the first three terms in \eqref{eqn:perihelionshift} (GR contributions) becomes
\begin{equation}
    \Delta\phi_{\rm GR, \rm Mercury}\approx 5.006 \cdot 10^{-7} \, \textrm{rad/cycles}  \approx 0.1033\, \textrm{''/cycles} \approx 42,84 \, \textrm{''/cen}\,.
\end{equation}
The observed value for the perihelion shift of Mercury is $42,98\pm 0.040\, \textrm{''/cen} $~\cite{Will:2005va,Will:1986ay}, so that, the maximum possible value is $43.02\, \textrm{''/cen}$ and the minimum is $42.94\, \textrm{''/cen}$. Since our perturbed solutions are only valid when $\epsilon\ll 1$, we then find the that the maximum value that $\Delta\phi_{\epsilon}$ could take for Mercury is
\begin{eqnarray}
 \Delta\phi_{\epsilon,\rm max}\approx 0.18 \, \textrm{''/cen}\,,
\end{eqnarray}
otherwise, the value of the perihelion shift will not be in the observed region. For example, for the Case 1a, we get that the maximum values for the constants must be
\begin{equation}
    \Big|\alpha +5.65\times 10^{7}\beta\Big|_{\rm max}\approx 3.65\times 10^{20}\, \textrm{km}^2\,.
\end{equation}
One can notice that for $\beta=0$, we get the same order of magnitude for $\alpha_{\rm max}\sim 10^{20}\, \textrm{km}^2$ as was found in Ref.~\cite{DeBenedictis:2016aze}. In our solutions, we have different constants, so that, each of them could have different maximum values. For example, if $\alpha=0$ (leading to a $B^2$ contribution), one finds that $\beta_{\rm max}\sim 10^{13}\, \textrm{km}^2$. Using the same idea for all the solutions, one gets that the constants must have the followings maximum expressions,
 \begin{eqnarray}
\textrm{ \textbf{Case 1a:}} &&\ \  \Big|\alpha+5.648\times 10^7 \beta\Big|_{\rm max} \approx 3.650\times 10^{20}\,\textrm{km}^2\,, \\ \nonumber\\
\textrm{ \textbf{Case 1b:}}&&\ \  \Big|\alpha+5.648\times 10^7 \beta+2.824\times 10^7 \gamma\Big|_{\rm max} \approx 3.650\times 10^{20}\,\textrm{km}^2\,, \\ \nonumber\\
\textrm{ \textbf{Case 1c:}}&&\ \  \Big|\alpha+5.648\times 10^7 \beta-1.207\times 10^{-22} \gamma \, \textrm{km}^{-2}\Big|_{\rm max} \approx3.650\times 10^{20}\,\textrm{km}^2\,, \\\nonumber \\
\textrm{ \textbf{Case 1d:}}&&\ \  \Big|\alpha+5.648\times 10^7 \beta-6.036\times 10^{-23} \gamma \, \textrm{km}^{-2}\Big|_{\rm max} \approx 3.650\times 10^{20}\,\textrm{km}^2\,, \\\nonumber \\
\textrm{ \textbf{Case 2a:}}&&\ \ \Big|- \zeta  \chi   (\xi+\chi)^2\Big|_{\rm max} \approx 2.018\times 10^{42}\,\textrm{km}^4\,, \\ \nonumber \\
\textrm{ \textbf{Case 2b:}}&&\  \Big|\zeta  \chi   (\xi+\chi)^3\Big|_{\rm max} \approx 1.402\times 10^{72}\,\textrm{km}^6\,.
\end{eqnarray}

For simplicity, we have set $\epsilon=1$ since this constant is just a tracking parameter for the expansion.

\subsection{Deflection of light}\label{light_deflec}
The deflection of light in strong fields provides a good tool to test  gravitational theories. This is especially interesting in the solar system where there is now numerous observations of light deflection events with which to compare. Since the Sun is the largest mass in the solar system, this normally takes place in the context of light rays being lensed by the Sun.

Let us now study how is the light deflection for the previously found perturbed solutions. We can define the minimal distance $r_0$ using $\dot{r}(r_0)=0$ ($V(r_0)=0$), and then using~\eqref{17}, we find
\begin{equation}
   r_0^2=\Big(\frac{h}{k}\Big)^2\mathcal{A}(r_0)\,,
\end{equation}
where we have set $\sigma=0$ since we are interested on masseless particles. By using the above equation and \eqref{potentialEq} and \eqref{kh} we get
\begin{equation}
    \frac{d\phi}{dr}=\frac{\dot{\phi}}{\dot{r}}=\pm\frac{h}{r^2\sqrt{-2V(r)}}=\pm\frac{\mathcal{B}^{1/2}}{r^2}\Big(\frac{\mathcal{A}(r_0)}{r_0^2\mathcal{A}}-\frac{1}{r^2}\Big)^{-1/2}\,.
\end{equation}
 Now, if we integrate this equation from a radius $r_0$ to $r$, we find the light deflection, namely
 \begin{equation}\label{phi}
    \phi(r)=\pm \int_{r_0}^r d\bar{r} \frac{\mathcal{B}(\bar{r})^{1/2}}{\bar{r}^2}\Big(\frac{\mathcal{A}(r_0)}{r_0^2\mathcal{A}(\bar{r})}-\frac{1}{\bar{r}^2}\Big)^{-1/2}\,.
\end{equation}
If we replace the metric functions and we expand up to first order in $\epsilon$, we find
\begin{eqnarray}\label{intphi}
\phi(r)= \pm r_0\int_{r_0}^r d\bar{r}\frac{ 4 M \left(r_0^3-\bar{r}^3\right)-2 r_0^3 \bar{r}+2 r_0 \bar{r}^3- \epsilon\,  r_0 \bar{r}^3 a(r_0)}{2 \bar{r}^2 \sqrt{\frac{2 M}{r_0}\left(\frac{r_0^3}{\bar{r}^3}-1\right)-\frac{r_0^2}{\bar{r}^2}+1} \left(2 M \left(r_0^3-\bar{r}^3\right)-r_0^3 \bar{r}+r_0 \bar{r}^3\right)}\,.
\end{eqnarray}
By choosing the positive sign in the integral, we find that the deviation angle is
\begin{eqnarray}
\vartheta(r)=2\phi(r)-\pi\,.\label{vartheta}
\end{eqnarray}
Now, we will replace $a(r)$ from each model, and assume that $r\gg 1$ and $r_0\gg1$. To do this, one needs to be careful with the expansions since for each model, the leading terms would appear at different order expansions in $r_0$. Since the models depend on the parameters $\alpha,\beta,\gamma$ and $\zeta$, we need to expand each solution only taking the leading terms for each parameter contribution. Since these expansions sometimes are difficult to understand, we will explicitly show the computations for the Case 1a. If one considers this case, the leading term for $r,r_0\gg 1$ in the integrand for the $\beta$ contribution is
\begin{eqnarray}\label{tildephibeta}
\tilde{\phi}_{\epsilon,\beta}(r_0,\bar{r})&\approx & \pm \epsilon\beta M^2 \sqrt{\frac{1}{\bar{r}^2-r_0^2}}\frac{4 r_0^4-r_0^2 \bar{r}^2-\bar{r}^4}{r_0^3 \bar{r}^5}+\mathcal{O}\Big(\bar{r}{}^{-7},r_0^{-5}\Big)\,,
\end{eqnarray}
and, the leading term in the integrand for the $\alpha$ contribution behaves as
\begin{eqnarray}\label{tildephialpha}
\tilde{\phi}_{\epsilon,\alpha}(r_0,\bar{r})&\approx & \pm 2\epsilon\alpha M^3 \sqrt{\frac{1}{\bar{r}^2-r_0^2}}\frac{5 r_0^6+5 r_0^5 \bar{r}+r_0^4 \bar{r}^2+r_0^3 \bar{r}^3+r_0^2 \bar{r}^4+r_0 \bar{r}^5+\bar{r}^6}{5 r_0^4 \bar{r}^6 (r_0+\bar{r})}+\mathcal{O}(\bar{r}{}^{-9},r_0^{-7})\,,
\end{eqnarray}
which clearly has  $r_0^{-6}$ and $\bar{r}^{-7}$ contributions which are not in the same order as the first leading contribution coming from $\beta$ (see Eq.~\eqref{tildephibeta}). The most important part is to always consider the leading contributions for each constant.

For the GR contribution, we will keep some extra terms to compare it with the $\epsilon$ contributions. The GR contribution in the integrand in \eqref{intphi} is approximated to (for $r,r_0\gg 1$)
\begin{eqnarray}\label{tildephiGR}
\tilde{\phi}(r_0,\bar{r})&\approx & M\sqrt{\frac{1}{\bar{r}^2-r_0^2}}\Big[\frac{r_0 }{M \bar{r}}+\frac{\left(r_0^2+r_0 \bar{r}+\bar{r}^2\right) }{\bar{r}^2 (r_0+\bar{r})}+\frac{3 M \left(r_0^2+r_0 \bar{r}+\bar{r}^2\right)^2 }{2 r_0 \bar{r}^3 (r_0+\bar{r})^2}\nonumber\\
&&+\frac{5 M^2 \left(r_0^2+r_0 \bar{r}+\bar{r}^2\right)^3 }{2 r_0^2 \bar{r}^4 (r_0+\bar{r})^3}\Big]+\mathcal{O}(\bar{r}{}^{-9},r_0^{-7})\,.
\end{eqnarray}
Then, for the Case 1a, if we assume $r\gg r_0$ and replace \eqref{tildephibeta} \eqref{tildephialpha} and \eqref{tildephiGR} into \eqref{phi}, and  then finally replace this expression in the deflection angle $\vartheta$ given in \eqref{vartheta}, one finds 
\begin{eqnarray}
    \vartheta &\approx& \frac{4 M}{r_0}+\frac{ M^2}{r_0^2}\left(\frac{15 \pi }{4}-4\right)+\frac{ M^3}{r_0^3}\Big(\frac{244-45 \pi}{6}\Big)+\epsilon  \frac{4 M^3 }{15 r_0^5} (16 \alpha +\beta )+\mathcal{O}\Big(r_0^{-7},\frac{r_0}{r}\Big)\,.
\end{eqnarray}
The first term is the standard GR deflection angle up to $r_0/r$ order, whereas the term multiplied by $\epsilon$ is the contribution from the Case 1a perturbed solution. As explained above, this final expansion also needs to be taken in a similar way, i.e., here we have expanded up to the leading term in $r_0/r$. In the Case 1a, the leading term is up to $r_0/r$, so $r$ does not appear. However, in other solutions, one would need to expand up to $(r_0/r)^{-2}$ to find the corresponding leading expansion terms. 

Following the same idea for all the other solutions, and expanding up to the leading corresponding appearing orders, we find that the deflection angle for all the models can be expressed as 
\begin{equation}
     \vartheta\approx \vartheta_{\rm GR}+\epsilon \, \vartheta_{\epsilon} = \frac{4 M}{r_0}+\frac{ M^2}{r_0^2}\left(\frac{15 \pi }{4}-4\right)+\frac{ M^3}{r_0^3}\Big(\frac{244-45 \pi}{6}\Big)+\epsilon\Big[ \frac{4 M^3 }{15 r_0^5} (16 \alpha +\beta )+ \gamma \vartheta_{\gamma}+\zeta\vartheta_{\zeta}\Big]\,,
\end{equation}
where for each case we have:
\begin{eqnarray}
\textrm{ \textbf{Case 1a:}}&&\ \ \gamma=\zeta=0\,,\label{light_case1a}\\
\textrm{ \textbf{Case 1b:}}&&\ \ \vartheta_{\gamma}\approx \frac{34   M^3  }{15 r_0^5}\,, \quad \zeta=0\,,\label{light_case1b}\\
 \textrm{ \textbf{Case 1c:}}&&\ \ \vartheta_{\gamma}\approx \frac{4 M^4}{r r_0^7}\,, \quad \zeta=0\,,\label{light_case1c}\\
\textrm{ \textbf{Case 1d:}} &&\ \ \vartheta_{\gamma}\approx \frac{2  M^4 }{r r_0^7}\,, \quad \zeta=0\,,\label{light_case1d}\\
 \textrm{ \textbf{Case 2a:}} &&\ \ 
\vartheta_{\zeta}\approx\frac{6  M^4 \chi  (\xi +\chi )^2}{r r_0^7}\,,\quad \alpha=\beta=\gamma=0\,,\label{light_case2a}\\
\textrm{ \textbf{Case 2b:}} &&\ \ 
\vartheta_{\zeta}\approx-\frac{16 \zeta  M^6 \chi    (\xi +\chi )^3}{r r_0^{11}}\,,\quad \alpha=\beta=\gamma=0\,.\label{light_case2b}
\end{eqnarray}

\noindent Now, one can use data from the Very Long Baseline Interferometry (VLBI) which uses radio-telescopes on Earth~\cite{robertson1991new}. Using the numerical values $M=M_{\odot}\approx 1.474\, \textrm{km}$, $r_0=2.35\times 10^{5}M_{\odot}$ and the distance of Earth from Sun $r=r_{\rm Earth}=5.08\times 10^{7}M_{\odot}$, one finds that the deflection of light in GR (near the Sun) becomes
\begin{eqnarray}
\vartheta_{\rm GR}\approx 8.511\times 10^{-6}\, \textrm{rad}\approx 1.756''\,.
\end{eqnarray}
It has been found that the observed measurements $\vartheta_{\rm obs}$ over the GR contribution $\vartheta_{\rm GR}$ is approximately constraint to be~\cite{Shapiro:2004zz,Will:2014kxa}
\begin{equation}
   \frac{\vartheta_{\rm obs}}{ \vartheta_{\rm GR}}\approx 1.0001\pm 0.0001\,,
\end{equation}
which gives that the $\epsilon$ correction being at most
\begin{equation}
    \vartheta_{\epsilon,\rm max}\approx 1.702\times 10^{-9}\,\textrm{rad}\approx 3.51\cdot 10^{-4}\,''\,.
\end{equation}
Using this maximum value, one then find for each solution that their maximum constants must be
 \begin{eqnarray}
\textrm{ \textbf{Case 1a:}}&&\ \Big|\alpha+6.25\times 10^{-2}\beta\Big|_{\rm max}\approx 1.998\times 10^{19}\, \textrm{km}^2\,, \\ \nonumber\\
\textrm{ \textbf{Case 1b:}}&&\ \Big|\alpha+6.25\times 10^{-2}\beta +5.313 \times 10^{-1}\gamma\Big|_{\rm max}\approx 1.998\times 10^{19}\, \textrm{km}^2\,, \\ \nonumber\\
\textrm{ \textbf{Case 1c:}}&&\ \Big|\alpha+6.25\times 10^{-2}\beta+1.923 \times 10^{-20} \gamma\,  \text{km}^{-2} \Big|_{\rm max}\approx 1.998\times 10^{19}\, \textrm{km}^2\,, \\\nonumber \\
\textrm{ \textbf{Case 1d:}}&&\ \Big|\alpha+6.25\times 10^{-2}+9.613\times 10^{-21} \gamma\, \text{km}^{-2} \Big|_{\rm max}\approx 1.998\times 10^{19}\, \textrm{km}^2\,, \\\nonumber \\
\textrm{ \textbf{Case 2a:}}&&\ \Big|\zeta  \chi    (\xi +\chi )^2 \Big|_{\rm max}\approx 6.983\times 10^{38} \, \textrm{km}^4\,, \\ \nonumber \\
\textrm{ \textbf{Case 2b:}}&&\ \Big|-\zeta  \chi   (\xi +\chi )^3 \Big|_{\rm max}\approx 2.740\times 10^{61}\, \textrm{km}^6\,.
\end{eqnarray}

\subsection{Cassini experiment}
It was found that the fractional frequency shift $y$ of a system composed of Earth-spacecraft-Earth (in a weak field limit) is given by~\cite{Bertotti:2003rm}
\begin{eqnarray}
y=2\frac{v_{\rm Cassini} l_{\rm Earth}+v_{\rm Earth}l_{\rm Cassini}}{l_{\rm Earth}+l_{\rm Cassini}}\vartheta\,,
\end{eqnarray}
where $\vartheta$ is the deflection angle for the light (found in the previous section), $l_{\rm Earth}$ and $l_{\rm Cassini}$ are  the distances from the Earth to the Sun and the Cassini spacecraft to the Sun, respectively, and $v_{\rm Earth}$ and $v_{\rm Cassini}$ are the transverse velocities of the of the Earth and the Cassini spacecraft, respectively. Now, if we assume that $l_{\rm Cassini}\gg l_{\rm Earth}$, we find that the GR contributions becomes
\begin{equation}
    y_{\rm GR}\approx 2\vartheta_{\rm GR}v_{\rm Earth}=\frac{8M}{r_0}v_{\rm Earth}\,.
\end{equation}
Since some $\vartheta_{\epsilon}$ contributions leading term depend on $r$, then, the $y_{\epsilon}$ expansion ($l_{\rm Cassini}\gg l_{\rm Earth}$) becomes
\begin{equation}
    y_{\epsilon}\approx v_{\rm Earth}(\vartheta(r_0,r_{\rm Earth})+\vartheta(r_0,r_{\rm Cassini}))\,.
\end{equation}
Then, for the general form of $y$ for each model can be written as
\begin{equation}
    y=y_{\rm GR}+\epsilon \,y_{\epsilon}\approx  \frac{8M}{r_0}v_{\rm Earth}+\epsilon\Big(\frac{8 M^3 (16 \alpha +\beta )}{15 r_{0}^5}+\gamma y_{\gamma}+\zeta y_{\zeta}\Big)v_{\rm Earth}\,,
\end{equation}
where for each model we have,
\begin{eqnarray}
\textrm{ \textbf{Case 1a:}}&&\ \ \gamma=\zeta=0\,,\label{cass_case1a}\\
\textrm{ \textbf{Case 1b:}}&&\ \ y_{\gamma}\approx \frac{68 M^3 }{15 r_{0}^5}\,, \quad \zeta=0\,,\label{cass_case1b}\\
 \textrm{ \textbf{Case 1c:}}&&\ \ y_{\gamma}\approx \frac{4   M^4  (r_{\rm Cassini}+r_{\rm Earth})}{r_{0}^7 r_{\rm Cassini} r_{\rm Earth}} \,, \quad \zeta=0\,,\label{cass_case1c}\\
\textrm{ \textbf{Case 1d:}} &&\ \ y_{\gamma}\approx \frac{2  M^4   (r_{\rm Cassini}+r_{\rm Earth})}{r_{0}^7 r_{\rm Cassini} r_{\rm Earth}}\,, \quad \zeta=0\,,\label{cass_case1d}\\
 \textrm{ \textbf{Case 2a:}} &&\ \ 
y_{\zeta}\approx \frac{6 M^4  \chi   (\xi +\chi )^2 (r_{\rm Cassini}+r_{\rm Earth})}{r_{0}^7 r_{\rm Cassini} r_{\rm Earth}}\,,\quad \alpha=\beta=\gamma=0\,,\label{cass_case2a}\\
\textrm{ \textbf{Case 2b:}} &&\ \ 
y_{\zeta}\approx -\frac{16  M^6  \chi   (\xi +\chi )^3 (r_{\rm Cassini}+r_{\rm Earth})}{r_{0}^{11} r_{\rm Cassini} r_{\rm Earth}}\,,\quad \alpha=\beta=\gamma=0\,.\label{cass_case2b}
\end{eqnarray}
The signal measured by Cassini is $y_{\rm obs}\sim 10^{-10}\pm 10^{-14}$~\cite{Bertotti:2003rm}. Moreover, the GR value can be found by replacing  $M_{\odot}=1.474 \, \textrm{km}$, $v_{\rm Earth}=9.93\times 10^{-5}$ (in units of the speed of light), $r_{\rm Cassini}=9.7\times 10^{8}M_{\odot}$ (distance of Saturn to the Sun), $r_0=4.7\times 10^{5}M_{\odot}, \ r_{\rm Earth}=1.016\times 10^{8} M_{\odot}$ (distance from the Earth to the Sun), giving
\begin{equation}
    y_{\rm GR}\approx 1.690\times 10^{-9}\,.
\end{equation}
Then, the approximated maximum values that one can get from this constrain is
 \begin{eqnarray}
\textrm{ \textbf{Case 1a:}}&&\ \Big|\alpha+6.25\times 10^{-2}\beta \Big|_{\rm max}\approx 1.053\times 10^{23}\, \textrm{km}^2\,, \\ \nonumber\\
\textrm{ \textbf{Case 1b:}}&&\ \Big|\alpha+6.25\times 10^{-2}\beta +5.313\times 10^{-1} \gamma\Big|_{\rm max} \approx  1.053\times 10^{23}\, \textrm{km}^2\,, \\ \nonumber\\
\textrm{ \textbf{Case 1c:}}&&\ \Big|\alpha+6.25\times 10^{-2}\beta+1.062 \times 10^{-20} \gamma\,  \text{km}^{-2} \Big|_{\rm max}\approx  1.053\times 10^{23}\, \textrm{km}^2\,, \\\nonumber \\
\textrm{ \textbf{Case 1d:}}&&\ \Big|\alpha+6.25\times 10^{-2}\beta+5.310\times 10^{-21} \gamma\, \text{km}^{-2}\Big|_{\rm max} \approx  1.053\times 10^{23}\, \textrm{km}^2\,, \\\nonumber \\
\textrm{ \textbf{Case 2a:}}&&\ \Big|\zeta  \chi    (\xi +\chi )^2\Big|_{\rm max}\approx 6.609\times 10^{42} \, \textrm{km}^4\,, \\ \nonumber \\
\textrm{ \textbf{Case 2b:}}&&\ \Big|-\zeta  \chi    (\xi +\chi )^3\Big|_{\rm max}\approx  2.627\times 10^{65}\, \textrm{km}^6\,.
\end{eqnarray}

\subsection{Retardation of light (Shapiro delay)}\label{shapiro_sec}
The Shapiro effect represents the time correction for the round trip of a radar signal that passes near a massive object in the presence of gravity \cite{PhysRevLett.13.789}. To find this, one needs to calculate the time required for a radial signal from two different points $r_0$ to $r$. By integrating Eq.~\eqref{potentialEq} from these two points we get that the time is
\begin{equation}\label{time}
    t(r,r_0)=\int\limits_{r_0}^r d\bar{r} \sqrt{-2V(\bar{r})}=\int\limits_{r_0}^r d\bar{r}\Big[\left(1-\frac{r_0^2 \mathcal{A}(r_0)}{\bar{r}^2 \mathcal{A}(\bar{r})}\right)\frac{\mathcal{A}(\bar{r})}{\mathcal{B}(\bar{r})}\Big]^{-1/2}\,,
\end{equation}
where we have used that $\dot{r}= 0$ at $r=r_0$, so $h^2/k^2=r_0^2/\mathcal{A}(r_0)$, $k=\dot{t}\mathcal{A}$ (see Eq.~\eqref{kh}) and $\sigma=0$ since we are dealing with photons. By replacing the metric functions and expanding up to first order in $\epsilon$, one finds that the integrand appearing in the above equation becomes
\begin{eqnarray}
\Big[\left(1-\frac{r_0^2 \mathcal{A}(r_0)}{\bar{r}^2 \mathcal{A}(\bar{r})}\right)\frac{\mathcal{A}(\bar{r})}{\mathcal{B}(\bar{r})}\Big]^{-1/2}&=&\frac{\mu_0 \left(\mu_0^2-1\right) }{\bar{\mu}^2}\Big(\mu_0^6-2 \mu_0^4+\mu_0^2-\bar{\mu}^2 \left(\bar{\mu}^2-1\right)^2\Big)^{-1/2}- \epsilon  \,\frac{\left(\mu_0^2-1\right)}{2 \mu_0 \bar{\mu}^4}\Big(\mu_0^6-2 \mu_0^4+\mu_0^2\nonumber\\
&&-\bar{\mu}^2 \left(\bar{\mu}^2-1\right)^2\Big)^{-3/2}\times \Big[\bar{\mu}^4 \left(\bar{\mu}^2-1\right)^2 a(r_0)+\mu_0^2 \left(\mu_0^6-2 \mu_0^4+\mu_0^2-2 \bar{\mu}^2 \left(\bar{\mu}^2-1\right)^2\right) a(\bar{r})\nonumber\\
&&-\mu_0^2 \bar{\mu}^4 \left(\mu_0^6-2 \mu_0^4+\mu_0^2-\bar{\mu}^2 \left(\bar{\mu}^2-1\right)^2\right) b(\bar{r})\Big]\,,\label{123}
\end{eqnarray}
where $\bar{\mu}^2=1-2M/\bar{r}$ and $\mu_0^2=1-2M/r_0$. If one integrates the GR contribution, one finds the standard time of light from $r_0$ to $r$ which is ($r,r_0\gg 1$)
\begin{equation}
    t_{\rm GR}(r,r_0)\approx M \sqrt{\frac{r-r_0}{r+r_0}}+2 M \log \left(\frac{\sqrt{r^2-r_0^2}+r}{r_0}\right)+\sqrt{r^2-r_0^2}+\mathcal{O}(M^2)\,.
\end{equation}
The so-called retardation of light (or Shapiro delay) is then defined as
\begin{equation}
    t_{\rm Shapiro}(r,r_0)=t(r,r_0)-\sqrt{r^2-r_0^2}\,,
\end{equation}
and the GR contribution is
\begin{equation}
     t_{\rm Shapiro, GR}(r,r_0)\approx 2M \sqrt{\frac{r-r_0}{r+r_0}}+4 M \log \left(\frac{\sqrt{r^2-r_0^2}+r}{r_0}\right)\approx -\frac{M r_0}{r}+2 M \log \left(\frac{2 r}{r_0}\right)+M+\mathcal{O}((r_0/r)^2)\,,
\end{equation}
where in the last expression we have approximated the term assuming $r\gg r_0$. For the $\epsilon$ contributions, one also need to integrate the expression~\eqref{123} after assuming $r,r_0\gg 1$ and expanding the terms. Under the extra condition $r\gg r_0$, one can then find that the time for all the solutions can be written as (considering only the leading terms for each constant contribution),
\begin{eqnarray}
     t_{\rm Shapiro}(r,r_0)&=& t_{\rm Shapiro, GR}(r_,r_0)+\epsilon\,  t_{ \rm Shapiro,\epsilon}(r_,r_0)\\
    &\approx& M \Big[1+2\log\Big(\frac{2r}{r_0}\Big)-\frac{r_0}{r}\Big]+\epsilon\Big[\frac{8 \alpha  M^3 }{3 r_0^4}-\frac{\beta  M^2  }{r r_0^2}+\gamma  \, t_{\rm Shapiro,\gamma}+\zeta \, t_{\rm Shapiro,\zeta}\Big]\,,
\end{eqnarray}
where for each solution we have
\begin{eqnarray}
\textrm{ \textbf{Case 1a:}}&&\ \ \gamma=\zeta=0\,,\label{shapiro_case1a}\\
\textrm{ \textbf{Case 1b:}}&&\ \ t_{\rm Shapiro,\gamma}\approx -\frac{ M^2  }{ 2r r_0^2} \,, \quad \zeta=0\,,\label{shapiro_case1b}\\
 \textrm{ \textbf{Case 1c:}}&&\ \ t_{\rm Shapiro,\gamma}\approx \frac{2  M^4}{r r_0^6} \  \,, \quad \zeta=0\,,\label{shapiro_case1c}\\
\textrm{ \textbf{Case 1d:}} &&\ \ t_{\rm Shapiro,\gamma}\approx \frac{ M^4 }{r r_0^6} \,, \quad \zeta=0\,,\label{shapiro_case1d}\\
 \textrm{ \textbf{Case 2a:}} &&\ \ 
t_{\rm Shapiro,\zeta}\approx \frac{3   M^4 \chi  (\xi +\chi )^2}{r r_0^6} \,,\quad \alpha=\beta=\gamma=0\,,\label{shapiro_case2a}\\
\textrm{ \textbf{Case 2b:}} &&\ \ 
t_{\rm Shapiro,\zeta}\approx  -\frac{8  M^6 \chi   (\xi +\chi )^3}{r r_0^{10}}\,,\quad \alpha=\beta=\gamma=0\,.\label{shapiro_case2b}
\end{eqnarray}
We can now use the observations from the  Viking
mission on Mars~\cite{Reasenberg:1979ey}. The delay in time for a radio signal that is emitted from Earth (orbital radius $r_{\rm Earth}$) to Mars (orbital radius $r_{\rm Mars}$) and back as the signal passes close to the Sun's surface $R\simeq R_{\odot}$, is given by
\begin{eqnarray}
    \Delta t &=& 2 \Big[t \left(r_{\rm Earth}, {R_{\odot}} \right) + t (r_{\rm Mars}, {R_{\odot}} ) - \sqrt{{r_{\rm Earth}}^2-{{R_{\odot}}^2}} - \sqrt{r^2_{\rm Mars}-{{R_{\odot}}^2}} \Big]\,,\\
    &\approx& 2\Big [t_{\rm Shapiro} \left(r_{\rm Earth}, {R_{\odot}} \right) +  t_{\rm Shapiro} (r_{\rm Mars}, {R_{\odot}} )\Big]  \,.
\end{eqnarray}
Let us now replace $r_{\rm Earth}\approx 1.016\times 10^8 M_{\odot}  ,\, r_{\rm Mars}\approx 1.542\times 10^8 M_{\odot}, \,R_{\odot}\approx 4.7\times 10^{5}M_{\odot}, \,  $ and $M=M_{\odot}\approx 1.474\, \textrm{km}$ for the above equation. For GR, this gives
\begin{equation}
     \Delta t_{\rm GR} 
     \approx 2.664\times 10^{-4}\, \textrm{s}\,,
\end{equation}
where we have change the value from the $c=1$ units to the standard units by diving the term by $c$. From observations, we know that~\cite{Reasenberg:1979ey}
\begin{equation}
  \frac{  \Delta t_{\rm obs}}{\Delta t_{\rm GR}}=1.000\pm 0.001\,,
\end{equation}
which gives the maximum correction from $\epsilon$ being
\begin{equation}
    \Delta t_{\epsilon, \rm max} \approx 8.880 \times 10^{-13}\, \textrm{s}\,. 
\end{equation}
Then, we can put some bound for all the solutions, that can be summarised as follows
 \begin{eqnarray}
\textrm{ \textbf{Case 1a:}}&&\ \Big|\alpha-6.763\times 10^{2}\beta \Big|_{\rm max}\approx 5.389\times 10^{20}\, \textrm{km}^2\,, \\ \nonumber\\
\textrm{ \textbf{Case 1b:}}&&\ \Big|\alpha-6.763\times 10^{2}\beta -3.381\times 10^{2} \gamma\Big|_{\rm max} \approx  1.053\times 10^{23}\, \textrm{km}^2\,, \\ \nonumber\\
\textrm{ \textbf{Case 1c:}}&&\ \Big|\alpha-6.763\times 10^{2}\beta +1.276\times 10^{-20} \gamma\,  \text{km}^{-2} \Big|_{\rm max}\approx  1.053\times 10^{23}\, \textrm{km}^2\,, \\\nonumber \\
\textrm{ \textbf{Case 1d:}}&&\ \Big|\alpha-6.763\times 10^{2}\beta +6.379\times 10^{-21}  \gamma\, \text{km}^{-2}\Big|_{\rm max} \approx  1.053\times 10^{23}\, \textrm{km}^2\,, \\\nonumber \\
\textrm{ \textbf{Case 2a:}}&&\ \Big|\zeta  \chi    (\xi +\chi )^2\Big|_{\rm max}\approx 2.816\times 10^{40} \, \textrm{km}^4\,, \\ \nonumber \\
\textrm{ \textbf{Case 2b:}}&&\ \Big|-\zeta  \chi    (\xi +\chi )^3\Big|_{\rm max}\approx  1.120\times 10^{63}\, \textrm{km}^6\,.
\end{eqnarray}

\subsection{Gravitational redshift and Equivalence Principle}
Let us suppose that light is propagating at different heights $r_1$ and $r_2$ ($r_1<r_2$). Then, the gravitational redshift is given by
\begin{eqnarray}
z\equiv\frac{\nu_2}{\nu_1}-1=\sqrt{\frac{\mathcal{A}(r_2)}{\mathcal{A}(r_1)}}-1\,,
\end{eqnarray}
where $\nu_1$ and $\nu_2$ are the frequencies measured from $r_1$ and $r_2$ respectively. Then, if one replaces the metric expanded around Schwarzschild~\eqref{AA}-\eqref{BB} and expand up to first order in $\epsilon$, one finds that the redshift is
\begin{eqnarray}
z= \frac{\nu_2}{\nu_1}-1\approx \frac{\mu_2}{\mu_1}+\epsilon  \left(\frac{a(r_1)}{2 \mu_1 \mu_2}-\frac{a(r_2) \mu_2}{2 \mu_1^3}\right)-1\,,
\end{eqnarray}
where $\mu_i(r)^2=1-2M/r_i$. Now, assuming that $\mu_2,\mu_1\gg 1$, and expanding up to the leading terms, we find that the GR contribution becomes
\begin{equation}
        \Big(\frac{\nu_2}{\nu_1}\Big)_{\rm GR}\approx 
        1+M(r_1^{-1}-r_2^{-1})\,.
\end{equation}
Now, for the $\epsilon$ corrections, we need to take the solutions (each model) and  assume $\mu_2,\mu_1\gg 1$. If we only consider the leading terms for each model (for each constant), we then find
\begin{equation}
\Big(\frac{\nu_2}{\nu_1}\Big)\approx \Big(\frac{\nu_2}{\nu_1}\Big)_{\rm GR}+\epsilon\Big[ \frac{2}{5}M^3\alpha  \left(r_1^{-5}-r_2^{-5}\right)+\beta M^2\left(r_1^{-4}-r_2^{-4}\right)+\gamma\Big(\frac{\nu_2}{\nu_1}\Big)_{\gamma}+\zeta\Big(\frac{\nu_2}{\nu_1}\Big)_{\zeta} \Big]\,,  
\end{equation}
where for each model we have
 \begin{eqnarray}
\textrm{ \textbf{Case 1a:}}&&\ \  \gamma=\zeta=0 \label{grav_red_case1a}\\  
\textrm{ \textbf{Case 1b:}}&&\ \ \Big(\frac{\nu_2}{\nu_1}\Big)_{\gamma}\approx  \frac{ M^2 }{2}\left(r_1^{-4}-r_2^{-4}\right) \,,\quad \zeta=0 \label{grav_red_case1b}\\  
\textrm{ \textbf{Case 1c:}}&&\ \ \Big(\frac{\nu_2}{\nu_1}\Big)_{\gamma}\approx   2M^4 \left(r_1^{-8}-r_2^{-8}\right) \,,\quad \zeta=0 \label{grav_red_case1c}\\  
\textrm{ \textbf{Case 1d:}}&&\ \ \Big(\frac{\nu_2}{\nu_1}\Big)_{\gamma}\approx  M^4\left(r_1^{-8}-r_2^{-8}\right) \,,\quad \zeta=0 \label{grav_red_case1d}\\ 
\textrm{ \textbf{Case 2a:}}&&\ \ \Big(\frac{\nu_2}{\nu_1}\Big)_{\zeta}\approx  3M^4 \xi (\xi+\chi)^2\left(r_1^{-8}-r_2^{-8}\right) \,,\quad \alpha=\beta=\gamma=0 \label{grav_red_case2a}\\  
\textrm{ \textbf{Case 2b:}}&&\ \ \Big(\frac{\nu_2}{\nu_1}\Big)_{\zeta}\approx  -8M^6 \xi (\xi+\chi)^3\left(r_1^{-12}-r_2^{-12}\right) \,,\quad \alpha=\beta=\gamma=0 \,.\label{grav_red_case2b}
\end{eqnarray}
To constrain our models, we can now use data from an experiment with a hydrogen-maser clock on a rocket launched to an altitude of about $10^{7}$ m~\cite{Vessot:1980zz}. Using this, we have that the observed value divided by the GR value takes the following value~\cite{Vessot:1980zz}
\begin{eqnarray}
\frac{\Delta \nu_{\rm obs}}{\Delta \nu_{\rm GR}}=1.000\pm0.0002\,,
\end{eqnarray}
where we have defined the difference in the frequencies as $\Delta \nu=\nu_2-\nu_1$. This would mean that the $\epsilon$ correction must have the following bound
\begin{eqnarray}
\frac{\Delta \nu_{\rm \epsilon}/\nu_2}{\Delta \nu_{\rm GR}/\nu_2}=\frac{\Big(\frac{\nu_2}{\nu_1}\Big)_{\epsilon}}{\Big(\frac{\nu_2}{\nu_1}\Big)_{\rm GR}-1}<2\times10^{-4}\,.
\end{eqnarray}
Now, let us take the data used to constrain the model. For this, one can set $r_1=1.436\times 10^{9}M_{\rm Earth}$ as the radius of the Earth and $r_2=3.68\times 10^9 M_{\rm Earth}$ as the distance from the experiment to the Earth (which is located at a height of $\sim 10^4$ km). Here, $M_{\rm Earth}\approx 4.426\times 10^{-6}\, \textrm{km}$ is the mass of the Earth in units $c=G=1$. By using these numbers in the above equation, we get that the solutions can have the following maximum value
\begin{equation}
    \Big|\Delta \nu_{\rm \epsilon}/\nu_2\Big|_{\rm max}\approx 8.492\times 10^{-14}\,,
\end{equation}
which for each model becomes
\begin{eqnarray}
\textrm{ \textbf{Case 1a:}}&&\ \Big|\alpha+3.539\times 10^9\beta \Big|_{\rm max}\approx 2.563\times 10^{22}\, \textrm{km}^2\,, \\ \nonumber\\
\textrm{ \textbf{Case 1b:}}&&\ \Big|\alpha+3.539\times 10^9\beta+1.769\times 10^9\gamma \Big|_{\rm max}\approx  2.563\times 10^{22}\, \textrm{km}^2\,, \\ \nonumber\\
\textrm{ \textbf{Case 1c:}}&&\ \Big|\alpha+3.539\times 10^9\beta-8.694\times 10^{-17}\gamma\, \textrm{km}^{-2}  \Big|_{\rm max}\approx  2.563\times 10^{22}\, \textrm{km}^2\,, \\\nonumber \\
\textrm{ \textbf{Case 1d:}}&&\ \Big|\alpha+3.539\times 10^9\beta-4.347\times 10^{-17}\gamma\, \textrm{km}^{-2}  \Big|_{\rm max}\approx  2.563\times 10^{22}\, \textrm{km}^2\,, \\\nonumber \\
\textrm{ \textbf{Case 2a:}}&&\ \Big|-\zeta  \chi    (\xi +\chi )^2 \Big|_{\rm max}\approx  1.965\times 10^{38} \, \textrm{km}^4\,, \\ \nonumber \\
\textrm{ \textbf{Case 2b:}}&&\ \Big|\zeta  \chi   (\xi +\chi )^3 \Big|_{\rm max}\approx  6.135\times 10^{63}\, \textrm{km}^6\,.
\end{eqnarray}

It is also clear that the potential in the Newtonian limit, $g_{00}\approx 1+2V$, depends only on $M$ and not on the mass or composition of the test particle for all the models. Then, the equivalence principle will not be violated in any of the models studied.

\section{Results and Conclusions }\label{sec:con}

$f(T,B)$ gravity represents a novel modification of GR that is not reproducible within the standard Levi-Civita connection framework of gravity since the scalars $T$ and $B$ represent a decoupling of the second and fourth order contributions of the Ricci scalar. This makes the $f(T,B)$ gravity scenario very interesting and may offer a new avenue of research for model proposals in $f(\lc{R})$ gravity.

In this work, we consider the scenario of a spherically symmetric metric ansatz in Eq.~(\ref{metric_ansatz}) from which we write the tetrad choice in Eq.~(\ref{tetrad_ansatz}). The field equations produced by $f(T,B)$ gravity must not only satisfy the regular ten linearly independent equations of motion related to the energy-momentum tensor, but they must also satisfy the vanishing six anti-symmetric equations of motion which represent the LLT invariance \cite{Li:2011wu}. In our case, we choose this particular tetrad because it leads to a zero spin connection components in Eq.~(\ref{FieldEq}). In the ensuing work, we choose a particular model for the arbitrary $f(T,B)$ Lagrangian in Eq.~(\ref{model_choice}) which leads to six separate weak field solutions. These six solutions represent the possible combinations of the free parameters of the model. An exact solution may again combine these weak field limits. However, for the purposes of weak field phenomenology, which is where Solar System test fall, this will suffice.

We first confront the issue of geodesics for both massless and massive particles for the weak fields of this model in subsection \ref{geodesics}. This leads immediately to the radial predictions for photon spheres which have recently become measurable \cite{Akiyama:2019cqa}, but more work needs to be done for this to become a constraining factor for theories beyond GR. This quantity is very interesting since it is related to the shadow of the black hole. We found that depending on the parameters, we have a larger or smaller value than the one predicted from GR. The inequalities leading to larger values are displayed in Eqs.~\eqref{sh_case1a}-\eqref{sh_case2b}. Another important feature of geodesics is the effect of perihelion shift which has been an observable astronomical ingredient for centuries. In Eqs.~(\ref{pw_case1a})--(\ref{pw_case2b}) we present leading order correction for each of the six cases for our weak field solution. An important property to identify is that in many of the cases, the free parameters appear in combination with each other which means that they may be made to be compatible with other phenomena such as cosmological tests. This is true for all the Solar System tests and may lead to more consistency in parameter fitting across tests in different scales of physics. Given that Mercury is the planet that had the largest expression of perihelion shift, we use observations of this effect on Mercury which naturally lead to constraints on parameter combinations which we summarize in Table~\ref{Table1}. In many cases, the constraints appear as a combination of model parameter which gives more freedom to render these models compatible with other tests.

Another crucial Solar System test is that of light deflection which has become more important in recent years due to its use in the H0licow result for the Hubble parameter at current times \cite{Wong:2019kwg}. In subsection~\ref{light_deflec}, we develop this for our six cases in which we study equatorial paths of light that are deflected by a mass $M$ resulting in the leading order predictions in excess to the GR result in Eqs.~(\ref{light_case1a})--(\ref{light_case2b}). In this case, we use light rays which are deflected by the Sun and which are measured by the VLBI which gives stronger constraints on the model parameters as compared with perihelion shift in some cases. The situation is made worse for the Cassini data which is retrieved by frequency delays due the round trip for signals being sent from Earth to the spacecraft. However, the general results in Eqs.~(\ref{cass_case1a})-(\ref{cass_case2b}) are applicable more generally for other situations of this kind.

In subsection \ref{shapiro_sec} we consider the case of Shapiro delay in which a radar signal is sent past the neighbourhood of a massive object such that the echo is slightly delayed due to the effect of gravity. Up to leading order, our six models predict Eqs.~(\ref{shapiro_case1a})-(\ref{shapiro_case2b}) in which again the model parameters appear in the same form as in the previous tests which makes direct comparison between the different solar system tests a more realistic undertaking. As can be seen in Table \ref{Table1}, the constraints on the model parameters are on par with the other tests. Finally, we explore the effect of gravitational redshift for the solution cases in Eqs.~(\ref{grav_red_case1a})-(\ref{grav_red_case2b}) which is constrained by signals being sent to Earth from a known source. The performance of this test is roughly equal to the other tests given the data on gravitational redshift.

In all the Solar System tests, the constraints are obtained by comparing the extra leading order terms produced by the particular phenomena against the analog GR term. This difference is then compared to observational data. Table \ref{Table1} gives a much clearer presentation of the results for each of the six cases of the weak field solution. As it shows, the constraints from each of the tests gives a roughly equal contribution to constraining the model parameters despite the differences in the precision of the tests which is an interesting feature of the of the results. It would be interesting to investigate other $f(T,B)$ gravity models and compare their constraints on model parameters. Another important avenue of research for this model would be to explore its cosmological consequences and whether these constraints can be refined using cosmological data.

\begin{table}[H]
\centering
 \begin{tabular}{|c|l |l | l | l | l|} 
\hline
  Model & Perihelion shift & Deflection Light & Cassini & Shapiro delay & Grav. redshift   \\ [0.5ex] \hline
Case 1a  
& $|\alpha+10^8 \beta|\lesssim  10^{20}$ & $|\alpha+10^{-1}\beta| \lesssim 10^{19}$& $\alpha+10^{-1}\beta \lesssim 10^{23}$
&$|\alpha-10^3\beta|\lesssim 10^{21}$
&$\alpha+10^9\beta \lesssim 10^{22}$\\ [0.5ex] \hline
Case 1b  
& $ |-\gamma|\lesssim  10^{13}$
& $|\gamma| \lesssim 10^{20}$
& $ |\gamma|\lesssim 10^{23}$ 
& $|-\gamma|\lesssim 10^{18}$
& $ | \gamma| \lesssim  10^{13}$
\\ [0.5ex] \hline
Case 1c  &
$|-\gamma| \lesssim  10^{42}$
&$|\gamma|\,  \lesssim 10^{39}$
&$| \gamma|\lesssim 10^{43}$
& $|\gamma|\lesssim 10^{40}$
&$|-\gamma| \lesssim 10^{38}$
\\ [0.5ex] \hline
Case 1d  & 
$| \gamma| \lesssim  10^{43}$ 
& $|\gamma|\lesssim 10^{39}$
& $ |\gamma|\lesssim 10^{43}$  
& $|\gamma|\lesssim 10^{41}$
& $|- \gamma|  \lesssim 10^{39}$
\\ [0.5ex] \hline
Case 2a  & 
$ |- \zeta  \chi   (\xi+\chi)^2|\lesssim  10^{42}$ & $|\zeta  \chi    (\xi +\chi )^2| \lesssim 10^{39} $
& $|\zeta  \chi    (\xi +\chi )^2| \lesssim  10^{43}$
&$|\zeta  \chi    (\xi +\chi )^2| \lesssim  10^{40}$
& $ |-\zeta  \chi    (\xi +\chi )^2| \lesssim  10^{38}$
\\ [0.5ex] \hline
Case 2b  & 
$| \zeta  \chi   (\xi+\chi)^3|\lesssim 10^{72}$ 
& $|-\zeta  \chi    (\xi +\chi )^3| \lesssim 10^{61}$
& $|-\zeta  \chi    (\xi +\chi )^3| \lesssim  10^{65}$
& $|-\zeta  \chi    (\xi +\chi )^2| \lesssim  10^{64}$
& $\ |\zeta  \chi    (\xi +\chi )^3| \lesssim  10^{64}$
\\ [0.5ex] \hline
 \end{tabular}
    \caption{constraints for the different solutions with different Solar System tests only considering the order of magnitudes of the maximum values of the parameters. The values have dimensions of $\textrm{km}^2,\, \textrm{km}^4$ and $\textrm{km}^6$ depending on the solutions, but we have omitted them here in order to safe space.  For each case, we have rounded the numbers to only show their order of magnitude.  Cases 1b-1d also contain the same $\alpha$ and $\beta$ contributions from Case 1a, but we have omitted them for simplicity to only show the order of magnitude in $\gamma$. These  contributions should also appear in Cases 1b-1d in the same way in Case 1a.} \label{Table1}
 \end{table}

As it was discussed in~\cite{DeBenedictis:2016aze}, even though the numbers in the table look large, they are not dimensionless quantity. Then, it may be made arbitrarily large or small by a simple change of units. One should clarify that if the order of magnitude appearing in this table is bigger, this would mean that the constant can take much bigger values, and hence, the contribution coming from that constant is much smaller and, that is why its contribution can take a much larger value. One notices that the $\gamma$ contribution highly depends on the solution. For the Cases 1c-1d with contributions $\gamma B T^2$ and $\gamma B^2 T$, respectively, one finds that their values are very small compared to the other contributions coming from the other modifications. For all Solar System tests, the leading terms for these solutions always appear at much higher order than the leading terms in $\alpha,\beta$ and $\gamma$ for the Case 1b with a contribution $\gamma B T$. For the Case 2, which behaves as $f(T,B)=T+ (1/2)\zeta (\chi T+\xi B)^u$, we notice that for a bigger $u$, the $\zeta$ contribution would produce much smaller values of each Solar System tests. This can also be seen in the leading terms in the expansions, i.e.,  when $u$ is larger, the leading terms in the expansions appear at a much higher order than the other solutions. Then, if $u$ is large enough, there will be almost not corrections to GR.

As a future work, it would be interesting to use the perturbed solutions found in this work for modelling galactic rotation curves as it was done in $f(T)$ gravity in~\cite{Finch:2018gkh}. The problem of this work is that they used an incorrect perturbed solution (around Minkowski) derived from a previous work~\cite{Ruggiero:2015oka}, where the authors used incorrect field equations in $f(T)$ since the tetrad chosen was not compatible with a vanishing spin connection. This problem can be easily seen in~\cite{Ruggiero:2015oka} since if $f(T)=T$, one does not recover the GR equations and then Schwarzschild as a solution. Moreover, if one uses the correct field equations with a tetrad compatible with the spin connection, the influence of the teleparallel perturbations for power-law $f(T)$ drops out for the Minkowski case (up to first order in $\epsilon$)~\cite{Bahamonde:2019zea}. Thus, it would be interesting to analyse what could happen with the correct perturbed solutions and also including the boundary term $B$ as we did here.

\section*{Acknowledgements}
The authors would like to acknowledge networking support by the COST Action GWverse CA16104. This article is based upon work from CANTATA COST (European Cooperation in Science and Technology) action CA15117, EU Framework Programme Horizon 2020. S.B. is supported by the European Regional Development Fund and the programm Mobilitas Pluss (Grants No. MOBJD423). ``M. Zubair thanks the Higher Education Commission, Islamabad, Pakistan for its
financial support under the NRPU project with grant number. The authors thank Christian Pfeifer for useful feedback.
$\text{7851/Balochistan/NRPU/R\&D/HEC/2017}$''. JLS would also like to acknowledge funding support from Cosmology@MALTA which is supported by the University of Malta.
\appendix
\section{Perturbed spherically symmetric equations}
\subsection{Perturbed equations}\label{appendix1}
If we assume a Schwarzschild background and a first order perturbation (See Eqs.~\eqref{AA} and \eqref{BB}) and then we take a power-law $f(T,B)$ as in Eq.~\eqref{model_choice}, we find that the field equations become ($\xi\neq-\chi$)
\begin{eqnarray}
&&\alpha  (-2)^{q+2} (\mu-1)^{-1}(q-1) r^{-2 q} \left(\frac{1}{\mu}-1\right)^{2 q} \mu^q \left(5 q \mu^2+2 q \mu+q+\mu-1\right)=\nonumber \\
&&\beta  (-1)^{m+1} 2^m r^{-2 m} (\mu-1)^{2 m-2} \mu^{-m-1} \Big[5 m \left(5 m^2-7 m+2\right) \mu^4+(m-1) m^2+4 \left(5 m^3-14 m^2+10 m-1\right) \mu^3\nonumber\\
&&+2 \left(7 m^3-20 m^2+9 m+4\right) \mu^2+4 (m-1)^3 \mu\Big]
 +(-1)^{s+w+1} 2^{s+w} \gamma (s+w-1) (\mu-1)^{2 s-2} r^{-2 (s+w)} \left(\frac{1}{\mu}-1\right)^{2 w} \times \nonumber\\
 &&\mu^{-s+w-1}\Big[5 \mu^4 \left(5 s^2+s (5 w-2)+4 w\right)+4 \mu^3 \left(5 s^2+s (5 w-9)-3 w+1\right)+2 \mu^2 \left(7 s^2+s (7 w-13)-2 (w+2)\right)\nonumber\\
&&+4 (s-1) \mu (s+w-1)+s (s+w)\Big]-\frac{16b_1 \mu^2 (\mu^2-2)}{r^2}+\frac{r^{1-2 u}}{r^2\mu(\xi +\chi ) (\mu-1)^2}\Big\{16 (\xi +\chi ) \mu^7 r^{2 u} b_1'\nonumber \\
&&-32 (\xi +\chi ) \mu^6 r^{2 u} b_1'+16 (\xi +\chi ) \mu^5 r^{2 u} b_1'+\zeta  r (-2)^{u+1} (u-1) (\mu-1)^{2 u} \mu^{2-u} (\xi +\chi )^u \Big[\left(7 u^2-13 u-4\right) \chi -2 \xi  (u+2)\Big]\nonumber\\
&&-5 \zeta  r (-2)^u (u-1) u (\mu-1)^{2 u} \mu^{4-u} (\xi +\chi )^u (4 \xi +(5 u-2) \chi )\Big)-\zeta  r (-2)^u (u-1) u^2 \chi  (\mu-1)^{2 u} \mu^{-u} (\xi +\chi )^u\nonumber\\
&&+\zeta  r (-1)^u 2^{u+2} (u-1)^2 (\mu-1)^{2 u} \mu^{1-u} (\xi +\chi )^u (\xi -u \chi +\chi )+\zeta  r (-1)^{u+1} 2^{u+2} (u-1) (\mu-1)^{2 u} \mu^{3-u} \times\nonumber\\
&&(\xi +\chi )^u \left(\xi +5 u^2 \chi -3 \xi  u-9 u \chi +\chi \right)\Big\}\,,\label{eqapp1}\\
&&b_1=\alpha  (-1)^q 2^{q-2} (q-1) r^{2-2 q} \left(\frac{1}{\mu}-1\right)^{2 q} \mu^{q-2}+a(\mu^2 -1)\mu^{-4}+(-1)^{m+1} 2^{m-4}\beta  r^{2-2 m} (\mu-1)^{2 m-2} \mu^{-m-3}\times\nonumber\\
&&\Big[2 \left(3 m^2-5 m+2\right) \mu^3+8 \left(m^2-1\right) \mu^2+2 \left(m^2-3 m+2\right) \mu+15 (m-1) m \mu^4+(m-1) m\Big]\nonumber\\
&&+(-1)^{s+w+1} 2^{s+w-4}\gamma   (s+w-1) (\mu-1)^{2 s-2}\mu^{-s+w-3} r^{-2 (s+w-1)} \left(\frac{1}{\mu}-1\right)^{2 w} \Big(15 s \mu^4+(6 s-4) \mu^3\nonumber\\
&&+8 (s+1) \mu^2+2 (s-2) \mu+s\Big) +\frac{r^{1-3 u}}{16 (\xi +\chi ) (\mu-1)^2 \mu^3}\Big[-2 \mu^3 \Big(8 (\xi +\chi ) r^{3 u} a'_1+(-1)^{u+1} 2^u \zeta  (u-1) r^{u+1}\times\nonumber\\
&&(\mu-1)^{2 u} \mu^{-u} (\xi +\chi )^u ((3 u-2) \chi -2 \xi )\Big)+8 \mu^2 \Big(4 (\xi +\chi ) r^{3 u} a'_1+ (-2)^u \zeta  (u-1) r^{u+1} (\mu-1)^{2 u} \mu^{-u} (\xi +\chi )^u \times\nonumber\\
&&(\xi +u \chi +\chi )\Big)+2 \mu \left(\zeta  (-2)^u (u-1) r^{u+1} (\mu-1)^{2 u} \mu^{-u} (\xi +\chi )^u ((u-2) \chi -2 \xi )-8 (\xi +\chi ) r^{3 u} a'_1\right)\nonumber\\
&&+15 \zeta  (-2)^u (u-1) u \chi  r^{u+1} (\mu-1)^{2 u} \mu^{4-u} (\xi +\chi )^u+\zeta  (-2)^u (u-1) u \chi  r^{u+1} (\mu-1)^{2 u} \mu^{-u} (\xi +\chi )^u\Big]\,.\label{eqapp2}
\end{eqnarray}
Note that the third field equation (Eq.~\eqref{Eq3}) is not an independent equation.
\subsection{Solutions for $f(T,B)=T+\frac{1}{2}\epsilon\, \zeta  (\xi  T+\chi B  )^u$, for $u=3$ and $u=4$}\label{appendix33}
The solution found for these cases are
\begin{itemize}
    \item $u=3$ yielding $T+\frac{1}{2}\epsilon\, \zeta  (\xi  T+\chi B  )^3$:\\
    \begin{eqnarray}
    \mathcal{A}(r)&=&\mu^2+\frac{\epsilon\zeta}{r^4 \left(\mu^2-1\right)^4}\Big[\frac{1}{315} (\xi +\chi )^2 (1787 \xi +2732 \chi )+9 \chi  (\xi +\chi )^2 \mu^{12}-\frac{4}{9} (\xi +\chi )^2 (2 \xi +83 \chi ) \mu^{11}\nonumber\\
    &&+3 (\xi +\chi )^2 (\xi +6 \chi ) \mu^{10}+\frac{8}{9} (\xi +\chi )^2 (4 \xi +139 \chi ) \mu^9-\frac{1}{3} (\xi +\chi )^2 (79 \xi +556 \chi ) \mu^8\nonumber\\
    &&+\frac{4}{35} (\xi +\chi )^2 (194 \xi -751 \chi ) \mu^7+\frac{1}{3} (\xi +\chi )^2 (172 \xi +1081 \chi ) \mu^6-\frac{16}{15} (\xi +\chi )^2 (112 \xi +157 \chi ) \mu^5\nonumber\\
    &&+(\xi +\chi )^2 (4 \xi -233 \chi ) \mu^4+\frac{4}{3} (\xi +\chi )^2 (206 \xi +305 \chi ) \mu^3-\frac{1}{45} (\xi +\chi )^2 (8987 \xi +5882 \chi ) \mu^2\nonumber\\
    &&-8 (\xi +\chi )^2 (4 \xi +13 \chi ) \mu-\frac{3 \chi  (\xi +\chi )^2}{\mu^2}+\frac{4 (\xi +\chi )^2 (2 \xi +5 \chi )}{\mu}+\log (\mu) \left(24 (\xi +\chi )^3-168 (\xi +\chi )^3 \mu^2\right)\Big]\,,\nonumber \\ \\
    \mathcal{B}(r)&=&\mu^{-2}+\frac{\epsilon\zeta}{r^4 \left(\mu^2-1\right)^4}\Big[-(\xi +\chi )^2 (64 \xi +367 \chi )+3 (\xi +\chi )^2 (4 \xi +13 \chi ) \mu^8-\frac{32}{9} (\xi +\chi )^2 (17 \xi +44 \chi ) \mu^7\nonumber\\
    &&+15 (\xi +\chi )^2 (5 \xi +6 \chi ) \mu^6+\frac{8}{63} (\xi +\chi )^2 (890 \xi +3347 \chi ) \mu^5-\frac{1}{3} (\xi +\chi )^2 (1037 \xi +1928 \chi ) \mu^4\nonumber\\
    &&+\frac{8}{35} (\xi +\chi )^2 (554 \xi -811 \chi ) \mu^3+\frac{13}{3} (\xi +\chi )^2 (92 \xi +209 \chi ) \mu^2-\frac{16}{15} (\xi +\chi )^2 (398 \xi +353 \chi ) \mu \nonumber\\
    &&+\frac{16 (\xi +\chi )^2 (52 \xi +61 \chi )}{3 \mu}-\frac{(\xi +\chi )^2 (33493 \xi +19318 \chi )}{315 \mu^2}-\frac{8 (2 \xi -\chi ) (\xi +\chi )^2}{\mu^3}\nonumber\\
    &&+\frac{(\xi +\chi )^2 (9553 \xi +8608 \chi )}{315 \mu^4}-\frac{8 (\xi +\chi )^2 (2 \xi +5 \chi )}{\mu^5}+\frac{9 \chi  (\xi +\chi )^2}{\mu^6}+\log (\mu) \left(\frac{24 (\xi +\chi )^3}{\mu^2}-\frac{24 (\xi +\chi )^3}{\mu^4}\right)\Big]\,.\nonumber \\ 
    \end{eqnarray}
     \item $u=4$ yielding $T+\frac{1}{2}\epsilon\, \zeta  (\xi  T+\chi B  )^4$:\\
    \begin{eqnarray}
    \mathcal{A}(r)&=&\mu^2+\frac{\epsilon\zeta}{r^6 \left(\mu^2-1\right)^6}\Big[-\frac{4 (7877 \xi -137983 \chi ) (\xi +\chi )^3}{2145}-24 \chi  (\xi +\chi )^3 \mu^{17}+\frac{12}{7} (\xi +\chi )^3 (\xi +85 \chi ) \mu^{16}\nonumber\\
    &&-\frac{16}{13} (\xi +\chi )^3 (8 \xi +177 \chi ) \mu^{15}+\frac{12}{7} (\xi +\chi )^3 (5 \xi -247 \chi ) \mu^{14}+\frac{24}{13} (\xi +\chi )^3 (32 \xi +851 \chi ) \mu^{13}\nonumber\\
    &&-\frac{32}{5} (\xi +\chi )^3 (23 \xi +93 \chi ) \mu^{12}-\frac{64}{33} (\xi +\chi )^3 (20 \xi +1703 \chi ) \mu^{11}+\frac{24}{5} (\xi +\chi )^3 (109 \xi +869 \chi ) \mu^{10}\nonumber\\
    &&-\frac{16}{21} (592 \xi -2705 \chi ) (\xi +\chi )^3 \mu^9-32 (\xi +\chi )^3 (22 \xi +223 \chi ) \mu^8+\frac{1056}{35} (\xi +\chi )^3 (48 \xi +83 \chi ) \mu^7\nonumber\\
    &&-8 (11 \xi -641 \chi ) (\xi +\chi )^3 \mu^6-\frac{48}{5} (\xi +\chi )^3 (208 \xi +553 \chi ) \mu^5+96 (\xi +\chi )^3 (19 \xi +\chi ) \mu^4\nonumber\\
    &&+64 (\xi +\chi )^3 (20 \xi +57 \chi ) \mu^3-\frac{8}{195} (\xi +\chi )^3 (25409 \xi +30089 \chi ) \mu^2-24 (\xi +\chi )^3 (32 \xi +61 \chi ) \mu\nonumber\\
    &&-\frac{12 (\xi +\chi )^3 (\xi +5 \chi )}{\mu^2}+\frac{8 \chi  (\xi +\chi )^3}{\mu^3}+\frac{16 (\xi +\chi )^3 (8 \xi +11 \chi )}{\mu}+\log (\mu) \left(192 (\xi +\chi )^4-2112 (\xi +\chi )^4 \mu^2\right)\Big]\,,\nonumber \\ \\
     \mathcal{B}(r)&=&\mu^{-2}+\frac{\epsilon\zeta}{r^6 \left(\mu^2-1\right)^6}\Big[96 (\xi +\chi )^3 (31 \xi -7 \chi )-16 (\xi +\chi )^3 (2 \xi +9 \chi ) \mu^{13}+\frac{12}{7} (\xi +\chi )^3 (131 \xi +495 \chi ) \mu^{12}\nonumber\\
    &&-\frac{16}{13} (\xi +\chi )^3 (394 \xi +1005 \chi ) \mu^{11}-\frac{4}{7} (\xi +\chi )^3 (365 \xi +3921 \chi ) \mu^{10}+\frac{96}{143} (\xi +\chi )^3 (3604 \xi +12041 \chi ) \mu^9\nonumber\\
    &&-\frac{32}{5} (\xi +\chi )^3 (427 \xi +477 \chi ) \mu^8-\frac{64}{33} (\xi +\chi )^3 (1454 \xi +7757 \chi ) \mu^7+\frac{24}{5} (\xi +\chi )^3 (1721 \xi +3761 \chi ) \mu^6\nonumber\\
    &&-\frac{32}{21} (1774 \xi -5177 \chi ) (\xi +\chi )^3 \mu^5-32 (\xi +\chi )^3 (275 \xi +802 \chi ) \mu^4+\frac{1056}{35} (\xi +\chi )^3 (302 \xi +267 \chi ) \mu^3\nonumber\\
    &&+8 (\xi +\chi )^3 (257 \xi +1789 \chi ) \mu^2-\frac{128}{5} (\xi +\chi )^3 (307 \xi +462 \chi ) \mu-\frac{8 (\xi +\chi )^3 (579871 \xi +494071 \chi )}{2145 \mu^2}\nonumber\\
    &&+\frac{48 (\xi +\chi )^3 (10 \xi +13 \chi )}{\mu^3}+\frac{4 (\xi +\chi )^3 (188057 \xi +42197 \chi )}{2145 \mu^4}-\frac{16 (\xi +\chi )^3 (18 \xi +25 \chi )}{\mu^5}\nonumber\\
    &&+\frac{36 (\xi +\chi )^3 (\xi +5 \chi )}{\mu^6}-\frac{32 \chi  (\xi +\chi )^3}{\mu^7}+\frac{64 (\xi +\chi )^3 (34 \xi +63 \chi )}{\mu}+\log (\mu) \left(\frac{192 (\xi +\chi )^4}{\mu^2}-\frac{192 (\xi +\chi )^4}{\mu^4}\right)\Big]\,.\nonumber \\
     \end{eqnarray}
\end{itemize}

\section{Photon sphere}\label{photonsphere}
For each solution we have that the photon sphere have the following contributions,
    \begin{eqnarray}
\textrm{ \textbf{Case 1a:}}\ \ \tilde{r}_1&=&0\,,\quad \tilde{h}_{1\pm}=0 \\  \textrm{ \textbf{Case 1b:}}\ \ \tilde{r}_1&=&\frac{\left(\sqrt{3}-33\right) \gamma }{9 \left(\sqrt{3}+3\right) M}+\frac{3 \gamma  \log (3)}{4 M}\approx \frac{0.0897709 \gamma }{M}\,,\\
\tilde{h}_{1\pm}&=&\pm \frac{2 \left(3 \sqrt{3}-5\right) \gamma  k_0}{\left(\sqrt{3}+3\right) M}\approx\pm \frac{0.0829038 \gamma  k_0}{M}\,,\\
\textrm{ \textbf{Case 1c:}}\ \ \tilde{r}_1&=&\frac{9 \gamma  \log (3)}{4 M^3}-\frac{\left(19909 \sqrt{3}+65263\right) \gamma }{8505 \left(\sqrt{3}+3\right) M^3}\approx-\frac{0.0065344 \gamma }{M^3}\,, \\
\tilde{h}_{1\pm}&=&\pm \frac{2 \left(14951 \sqrt{3}-125323\right) \gamma  k_0}{8505 \left(\sqrt{3}+1\right) M^3}\pm\frac{9 \left(\sqrt{3}+3\right) \gamma  k_0 \log (3)}{2 \left(\sqrt{3}+1\right) M^3}\approx\pm\frac{0.00484653 \gamma  k_0}{M^3}\,,\\ \textrm{\textbf{ Case 1d:}}\ \ \tilde{r}_1&=&\frac{9 \gamma  \log (3)}{4 M^3}-\frac{\left(17809 \sqrt{3}+69043\right) \gamma }{8505 \left(\sqrt{3}+3\right) M^3}\approx-\frac{0.0100799 \gamma }{M^3}\,, \\
\tilde{h}_{1\pm}&=&\pm\frac{2 \left(14951 \sqrt{3}-125323\right) \gamma  k_0}{8505 \left(\sqrt{3}+1\right) M^3}\pm\frac{9 \left(\sqrt{3}+3\right) \gamma  k_0 \log (3)}{2 \left(\sqrt{3}+1\right) M^3}\approx \pm \frac{0.00484653 \gamma  k_0}{M^3}\,,\\
 \textrm{ \textbf{Case 2a:}}\ \ \tilde{r}_1&=&\frac{9 \zeta  \log (3) (\xi +\chi )^3}{4 M^3}-\frac{16 \zeta  (\xi +\chi )^2 \left(\left(15709 \sqrt{3}+72823\right) \xi +\left(22009 \sqrt{3}+61483\right) \chi \right)}{105 \left(\sqrt{3}-3\right)^4 \left(\sqrt{3}+3\right)^5 M^3}\nonumber\\
 &\approx&-\frac{0.0136 \zeta  (\xi +0.219 \chi ) (\xi +\chi )^2}{M^3}\,,\\
 \tilde{h}_{1\pm}&=&\pm \zeta  \frac{k_0}{M^3} \left(\frac{9}{2} \sqrt{3} \log (3)-\frac{2 \left(70137 \sqrt{3}-85088\right)}{8505}\right) (\xi +\chi )^3 \nonumber\\
 &\approx& \pm \frac{0.00485 \zeta  k_0 (\xi +\chi )^3}{M^3} \,, \quad \textrm{with}\ \  \alpha=\beta=0\,,\\
 \textrm{\textbf{Case 2b:}}\ \ \tilde{r}_1&=&\frac{9 \zeta  \log (3) (\xi +\chi )^4}{2 M^5}-\frac{2 \zeta  (\xi +\chi )^3 \left(\left(16860416 \sqrt{3}-2152725\right) \xi +3 \left(7008192 \sqrt{3}-3119975\right) \chi \right)}{10945935 M^5}\nonumber\\
 &\approx& \frac{0.001213 \zeta  (\xi +0.1962 \chi ) (\xi +\chi )^3}{M^5}\,,\\ \tilde{h}_{1\pm}&=&\pm\frac{2 \zeta  k_0 \left(5610368 \sqrt{3}+26358900-32837805 \log (3)\right) (\xi +\chi )^4}{1216215 \sqrt{3} M^5}\nonumber\\
 &\approx&\pm \frac{0.0003098 \zeta  k_0 (\xi +\chi )^4}{M^5}\,, \quad \textrm{with}\ \  \alpha=\beta=0\,.
\end{eqnarray}

\bibliographystyle{utphys}
\bibliography{references}

\end{document}